\documentclass[journal]{IEEEtran}

\usepackage{soul}
\usepackage{wrapfig}
\usepackage{booktabs}
\usepackage{algorithm}
\usepackage{algorithmic}
\usepackage{amsmath}
\usepackage[table]{xcolor} % xcolor + colortbl 功能
\usepackage{colortbl} 
\usepackage{caption}
\usepackage{pifont}
\usepackage{multirow}
\usepackage{makecell}
\usepackage[colorlinks, citecolor=blue]{hyperref}
\definecolor{deepblue}{rgb}{0, 0, 0.9}
\definecolor{lightgray}{rgb}{0.4, 0.4, 0.4}
\usepackage{arydshln}
\usepackage{graphicx}
\usepackage{wrapfig}
\usepackage{subcaption}
\usepackage{url}            % simple URL typesetting
\usepackage{amsfonts}       % blackboard math symbols
\usepackage{nicefrac}       % compact symbols for 1/2, etc.
\usepackage{microtype}      % microtypography
\usepackage{mathabx}
\usepackage{hyperref}
\usepackage{graphicx}
\begin{document}
\title{Mamba in Speech: Towards an Alternative to Self-Attention}
\author{Xiangyu Zhang\textsuperscript{†}, 
Qiquan Zhang\textsuperscript{†}~\IEEEmembership{Member,~IEEE,} 
Hexin Liu\textsuperscript{†}~\IEEEmembership{Member,~IEEE,} 
Tianyi Xiao, 
Xinyuan Qian ~\IEEEmembership{Senior Member,~IEEE,} 
Beena Ahmed~\IEEEmembership{Member,~IEEE,} 
Eliathamby Ambikairajah~\IEEEmembership{Life Senior Member,~IEEE,} \\
Haizhou Li~\IEEEmembership{Fellow,~IEEE}, 
Julien Epps~\IEEEmembership{Senior Member,~IEEE}

\thanks{\textsuperscript{†}Equal contribution}
\thanks{\textcolor{black}{Xiangyu Zhang, Beena Ahmed, and Julien Epps are supported by the Australian Research Council Discovery Project DP230101184. Qiquan Zhang and Haizhou Li are supported in part by Shenzhen Science and Technology Program (Shenzhen Key Laboratory, Grant No. ZDSYS20230626091302006), and in part by Shenzhen Science and Technology Research Fund (Fundamental Research Key Project, Grant No. JCYJ20220818103001002). Haizhou Li is also supported by Program for Guangdong Introducing Innovative and Enterpreneurial Teams, Grant No. 2023ZT10X044. \textit{(Corresponding author: Qiquan Zhang; Hexin Liu)}}}
% \thanks{This work is supported by \textcolor{red}{Funding information} \textcolor{blue}{\textit{(Corresponding Authors: Qiquan Zhang; Hexin Liu)}}}
\thanks{Xiangyu Zhang, Qiquan Zhang, Beena Ahmed, Eliathamby Ambikairajah, and Julien Epps are with the School of Electrical Engineering and Telecommunications, The University of New South Wales (UNSW), Sydney, Australia \textcolor{black}{(e-mail: {xiangyu.zhang2}@unsw.edu.au; {zhang.qiquan}@outlook.com; \{beena.ahmed; e.ambikairajah; j.epps\}@unsw.edu.au).}}
\thanks{Hexin Liu, Tianyi Xiao are with the College of Computing and Data Science, Nanyang Technological University, Singapore \textcolor{black}{(e-mail: hexin.liu@ntu.edu.sg)}.}
\thanks{\textcolor{black}{Xinyuan Qian is with the School of Computer and Communication Engineering, University of Science and Technology Beijing, Beijing, 100083, China (e-mail: qianxy@ustb.edu.cn).}}
\thanks{\textcolor{black}{Haizhou Li is with the Guangdong Provincial Key Laboratory of Big Data Computing, The Chinese University of Hong Kong (Shenzhen), 518172 China, and also with Shenzhen Research Institute of Big data, Shenzhen, 51872 China (e-mail: {haizhouli}@cuhk.edu.cn).}}
\thanks{The code implementation can be found in \url{https://github.com/Tonyyouyou/Mamba-in-Speech}}
}

% % The paper headers
% \markboth{IEEE TRANSACTIONS ON MICROWAVE THEORY AND TECHNIQUES, VOL.~60, NO.~12, DECEMBER~2012
% }{Roberg \MakeLowercase{\textit{et al.}}: High-Efficiency Diode and Transistor Rectifiers}

% ====================================================================
\maketitle

\begin{abstract}
Transformer and its derivatives have achieved success in diverse tasks across computer vision, natural language processing, and speech processing. To reduce the complexity of computations within the multi-head self-attention mechanism in Transformer, Selective State Space Models (i.e., Mamba) were proposed as an alternative. Mamba exhibited its effectiveness in natural language processing and computer vision tasks, but its superiority has rarely been investigated in speech signal processing. This paper explores solutions for applying Mamba to speech processing by discussing two typical speech processing tasks: speech recognition, which requires semantic and sequential information, and speech enhancement, which focuses primarily on sequential patterns. The experimental results confirm that bidirectional Mamba (BiMamba) consistently outperforms vanilla Mamba, highlighting the advantages of a bidirectional design for speech processing. Moreover, experiments demonstrate the effectiveness of BiMamba as an alternative to the self-attention module in the Transformer model and its derivates, particularly for the semantic-aware task. The crucial technologies for transferring Mamba to speech are then summarized in ablation studies and the discussion section, offering insights for extending this research to a broader scope of tasks. 

\end{abstract}
\maketitle
\begin{IEEEkeywords}
Speech Recognition, Speech Enhancement, Mamba, Selective State Space Model, Self-Attention
\end{IEEEkeywords}

\IEEEpeerreviewmaketitle
\section{Introduction}

Transformer-based models~\cite{vaswani2017attention} have shone brightly across various domains in machine learning, including computer vision (CV)~\cite{dosovitskiy2020image,arnab2021vivit,liu2021swin}, natural language processing (NLP)~\cite{kenton2019bert,liu2023gpt,touvron2023llama}, and speech processing~\cite{speech_transformer, gulati20_interspeech, peng2022branchformer, chen2022wavlm}.  
This success is linked to the multi-head self-attention~(MHSA) module, which facilitates the representation of intricate data structures within a specific context window. However, the self-attention mechanism encounters a challenge with computational complexity, which grows quadratically as the size of the context window increases. In speech tasks, the window typically encompasses an entire speech signal, leading to substantial computational demands, particularly for frame-level acoustic feature sequences.

To address this challenge, state space models (SSMs) have emerged as a promising alternative~\cite{gu2021combining, gu2021efficiently, gupta2022diagonal, gu2022parameterization}. By integrating a time-varying mechanism into SSMs, a selective SSM named Mamba~\cite{gu2023mamba} has been proposed and shown outstanding performance to be effective in CV~\cite{zhu2024vision,yang2024vivim} and NLP~\cite{lieber2024jamba}. However, in the field of speech processing, despite some attempts to replace Transformers with Mamba~\cite{jiang2024dual,li2024spmamba,quan2024multichannel}, the results have not been as satisfactory as expected. In~\cite{jiang2024dual}, each Mamba is directly employed as a substitute for a Transformer within a dual-path framework for speech separation. \cite{li2024spmamba} proposed SPMamba for speech separation, where Mamba is used in conjunction with MHSA. Although these approaches achieve high performance by employing a dual-path strategy or combining Mamba with attention to form a new module, these methods negate the low time complexity of Mamba. In the domain of multi-channel speech enhancement, Mamba was implemented to enhance a SpatialNet from offline to online~\cite{quan2024multichannel} yet underperformed the vanilla version.

Since different speech tasks focus on various characteristics of a speech signal (e.g., speaker, language, emotion), they generally require different levels of information. However, existing approaches have mostly investigated speech enhancement and separation tasks, which focus primarily on the low-level information within a speech signal. Therefore, it is still unclear how to efficiently employ Mamba for other speech tasks, such as speech recognition and spoken language understanding, which require high-level semantic information within the speech signals.

\begin{figure*}[t]
    \hspace{2em}
    \begin{subfigure}[b]{0.40\textwidth}
        \centering
        \includegraphics[width=0.7\linewidth]{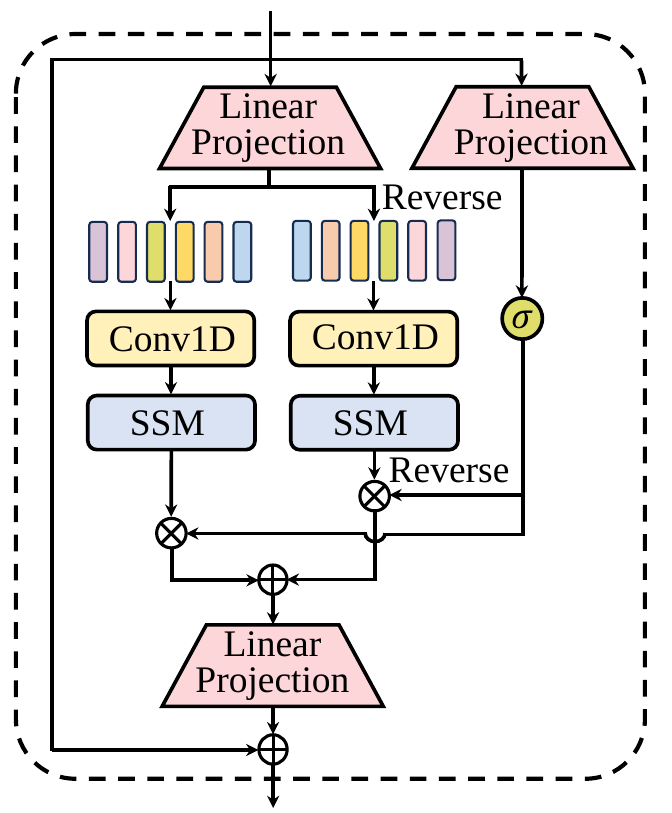}
        \caption{InnBiMamba}
        \label{inner}
    \end{subfigure}
    \hfill 
    \begin{subfigure}[b]{0.42\textwidth}
        \centering
        \includegraphics[width=\linewidth]{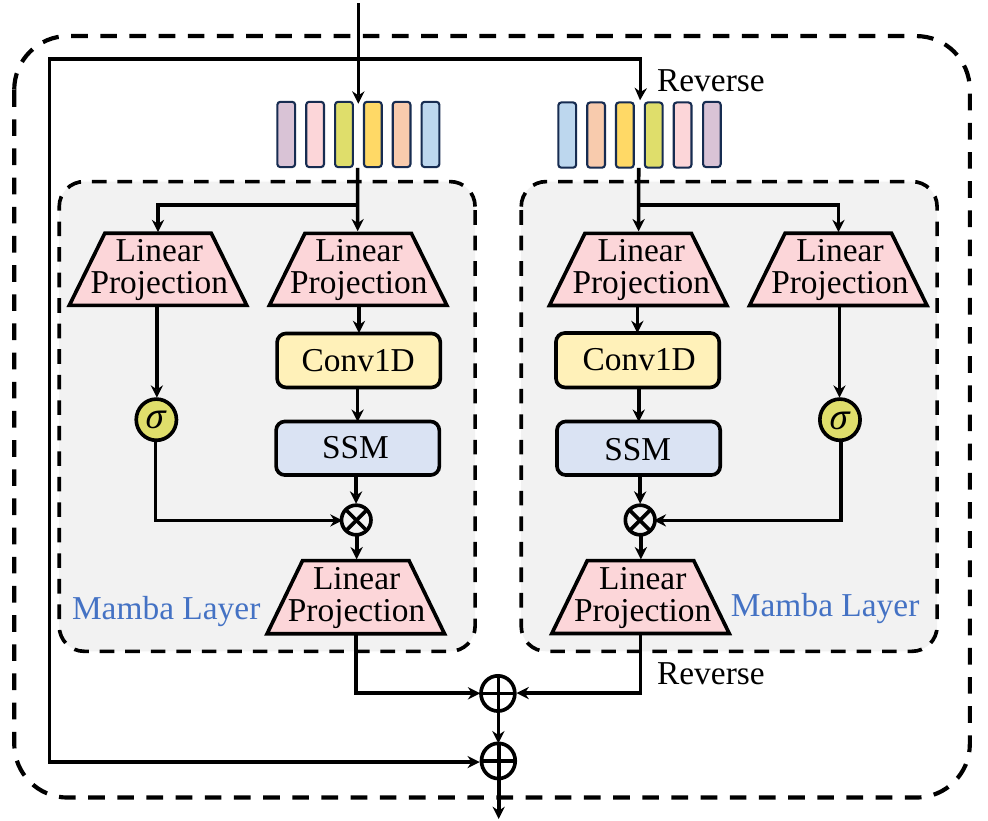}
        \caption{ExtBiMamba}
        \label{OuteVr}
    \end{subfigure}
    \hspace{2em}
\caption{The illustrations of (a) inner bidirectional Mamba (InnBiMamba) from Vison Mamba~\cite{zhu2024vision}, and (b) the external bidirectional Mamba (ExtBiMamba), where $\sigma$ denotes the SiLU activation.}
\label{Bi-directional}
\vspace{-1.0em}
\end{figure*}
In this paper, we explore solutions for applying Mamba to different speech tasks based on their varying information requirements (in different abstraction levels~\cite{li2013spoken}) using speech recognition and speech enhancement as examples. We first introduce and compare two bidirectional Mamba~(BiMamba) structures, external BiMamba (ExtBiMamba) and inner BiMamba (InnBiMamba), in speech tasks. The experiments confirm that a bidirectional design can enhance the capability of Mamba to model global dependencies within the features of a speech signal. Mamba and BiMamba models are then evaluated independently or as replacements for MHSA in Transformer and Conformer models. We demonstrate that the proposed BiMamba modules require additional nonlinearity to effectively learn high-level semantic information in speech tasks. Therefore, using BiMamba as an alternative to MHSA presents an optimal approach for applying Mamba in this scenario.

\section{Preliminary}
\label{sec:preliminary}

\subsection{State Space Models and Mamba}  

The State Space Model (SSM) based architectures, such as the Structured State Space Sequence (S4) Model~\cite{gu2021efficiently} and Mamba~\cite{gu2023mamba}, are typically inspired by the continuous linear time-invariant (LTI) systems. It maps a sequence $x(t)\in\mathbb{R}$ to $y(t)\in\mathbb{R}$ by leveraging a hidden state $\mathbf{h}(t)\in\mathbb{R}^{N\times 1}$. Formally, the mapping process of SSMs can be formulated as follows:
\begin{equation}\label{continue_equation}
\begin{aligned}
    \mathbf{h}'(t) &= \mathbf{A}\mathbf{h}(t) + \mathbf{B}x(t), \\
    y(t) &= \mathbf{C}\mathbf{h}'(t) + Dx(t),
\end{aligned}
\end{equation}
where $\mathbf{A}\!\in\!\mathbb{R}^{N\times N}$, $\mathbf{B}\!\in\!\mathbb{R}^{N\times 1}$, $\mathbf{C}\!\in\!\mathbb{R}^{1\times N}$, and $D\!\in\!\mathbb{R}$ (optional) represent the continuous SSM parameters~\cite{gu2023mamba}.

\textbf{SSM Discretization.} In Mamba and S4, a discretization process is applied to continuous-time SSMs for the integration into practical deep neural architectures. A timescale parameter $\Delta\in\mathbb{R}$ is introduced to transform the continuous matrices $\mathbf{A}, \mathbf{B}$ to their discrete counterparts $\overline{\mathbf{A}}, \overline{\mathbf{B}}$. The zero-order hold (ZOH) is a commonly used approach for the transformation, which is formulated as follows:
\begin{equation}
\begin{aligned}
    \overline{\mathbf{A}} &= \exp{(\Delta\mathbf{A})}, \\
    \overline{\mathbf{B}} &= (\Delta \mathbf{A})^{-1} (\exp{\Delta} \mathbf{A} - \mathbf{I})\cdot \Delta \mathbf{B} \\
    \end{aligned}
\end{equation}
The approximation of $\overline{\mathbf{B}}$ refined using the first-order Taylor series is given by $\overline{\mathbf{B}}\approx (\Delta \mathbf{A})(\Delta \mathbf{A})^{-1} \Delta \mathbf{B} \approx \Delta \mathbf{B}$. Thus, the discrete SSM is expressed as:
\begin{equation}
\begin{aligned}
    \mathbf{h}'(t) & = \overline{\mathbf{A}}\mathbf{h}(t) + \Delta{\mathbf{B}}x(t), \\
    y(t) &= \mathbf{C}\mathbf{h}'(t) + Dx(t).
\end{aligned}
\end{equation}
The output sequence can be computed simultaneously through a global convolution operation as follows:
\begin{equation}
\begin{aligned}
\overline{\mathbf{K}} & =\left(\mathbf{C} \overline{\mathbf{B}}, \mathbf{C} \overline{\mathbf{A B}}, \ldots, \mathbf{C} \overline{\mathbf{A}}^{L-1} \overline{\mathbf{B}}, \ldots\right) \\
\mathbf{y} & =\mathbf{x} \circledast \overline{\mathbf{K}},
\end{aligned}
\end{equation}
where $\overline{\mathbf{K}}$ is a convolution kernel derived from the SSM, $L$ is the length of the input sequence, and $\circledast$ denotes the convolution operation. Since the parameters $\mathbf{A}$, $\mathbf{B}$, and $\mathbf{C}$ remain constant with respect to input contents (temporal dynamics), previous S4 models struggle to effectively capture contextual information.

\textbf{Selective SSM.} Mamba~\cite{gu2023mamba} improves S4 by incorporating a selective mechanism that enables adaptive parameter adjustment based on input characteristics, presenting the selective state space model. The parameters $\mathbf{\Delta}$, $\mathbf{B}$, and $\mathbf{C}$ are computed as functions of the input. Given the input sequence $\mathbf{X} = \{\boldsymbol{x}_l\,|\,\boldsymbol{x}_l \in \mathbb{R}^{1\times E}, l=1,..., L\}$ and the output sequence $\mathbf{Y} = \{\boldsymbol{y}_l\,|\,\boldsymbol{y}_l \in \mathbb{R}^{1\times E}, l=1,..., L\}$, the selective SSM handles each channel independently through the hidden state $\mathbf{h}_{l}\in \mathbb{R}^{N \times E}$, and the formulation can be written as follows~\cite{han2024demystify}:
\begin{equation}\label{mamba_euqation}
\begin{aligned}
    \mathbf{h}_{l} & = \overline{\mathbf{A}}_{l} \odot \mathbf{h}_{l-1} +\mathbf{B}_{l}(\mathbf{\Delta}_{l}\odot \boldsymbol{x}_{l}), &\,\,\, \overline{\mathbf{A}}_{l} \in \mathbb{R}^{N \times E} \\
    \boldsymbol{y}_{l} &= \mathbf{C}_{l}\mathbf{h}_{l} + \mathbf{D}\odot \boldsymbol{x}_{l}, &\,\,\, \mathbf{D}\in\mathbb{R}^{1\times E}, \\
\end{aligned}
\end{equation}
where $\mathbf{B}_{l}\in\mathbb{R}^{N\times 1}$, $\mathbf{C}_{l}\in\mathbb{R}^{1 \times N}$, and $\mathbf{\Delta}_{l}\in\mathbb{R}^{1\times E}$ are derived from the input, and $\odot$ denotes the Hadamard product. Mamba generates the parameters $\mathbf{B}\in\mathbb{R}^{N \times L}$, $\mathbf{C}\in\mathbb{R}^{L \times N}$, and $\mathbf{\Delta}\in\mathbb{R}^{L \times E}$ using the following projections:
$\mathbf{B} = (\mathbf{X}\mathbf{W}_B)^\top$, $\mathbf{C} = \mathbf{X}\mathbf{W}_C$, $ \mathbf{\Delta} = \text{Softplus}(\mathbf{X}\mathbf{W}_1 \mathbf{W}_2)$ where \( \mathbf{W}_B, \mathbf{W}_C \in \mathbb{R}^{E \times N} \), \( \mathbf{W}_1 \in \mathbb{R}^{E \times \lceil E/r\rceil }\), and \( \mathbf{W}_2 \in \mathbb{R}^{\lceil E/r \rceil \times E} \) are projection matrices, and $\text{softplus}(\cdot)$ refers to $\log\bigl(1 + \exp(\cdot)\bigr)$. The reduction ratio $r$ of the two-layer linear projection for $\mathbf{\Delta}$ is set to 16 by default~\cite{gu2023mamba}. Equation~\ref{mamba_euqation} represents the exact selective SSM used in Mamba, with only slight formatting adaptations.

\begin{figure*}[!htbp]
    \begin{subfigure}[b]{0.3\textwidth}
        \centering
        \includegraphics[width=0.81\linewidth]{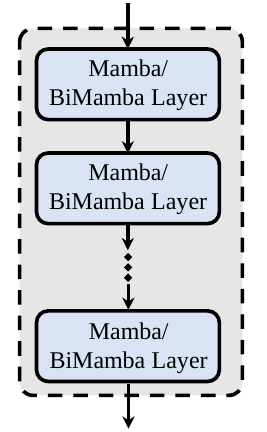}
        \caption{Mamba model}
        \label{Pure Mamba}
    \end{subfigure}
    \hfill 
    \begin{subfigure}[b]{0.31\textwidth}
        \centering
        \includegraphics[width=0.7\linewidth]{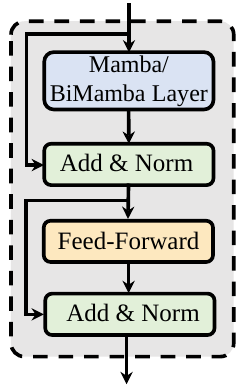}
        \caption{Transformer layer with Mamba}
        \label{Transformer-Mamba}
    \end{subfigure}
    \hfill 
    \begin{subfigure}[b]{0.30\textwidth}
        \centering
        \includegraphics[width=0.65\linewidth]{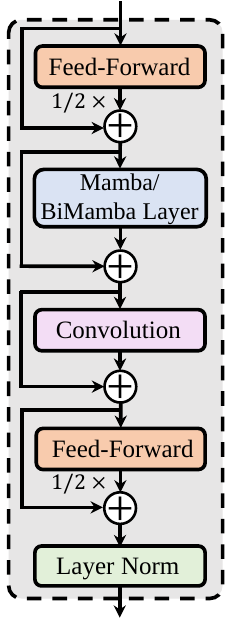}
        \caption{Conformer layer with Mamba}
        \label{Con-Mamba}
    \end{subfigure}

     \caption{Three applications of the Mamba layer in speech processing include: (a) using stacked unidirectional/bidirectional Mamba layers as an alternative to Transformer layers; (b) replacing causal and non-causal MHSA in Transformer layer with unidirectional/bidirectional Mamba, \textcolor{black}{termed TransMamba and TransBiMamba}; and (c) \textcolor{black}{replacing MHSA in Conformer layer with Mamba, \textcolor{black}{termed ConMamba and ConBiMamba.}}}
    \label{Methodology}
    \vspace{-1.0em}
\end{figure*}
The selective mechanism features the dynamic parameters, allowing the model to learn context-aware representations and effectively filter out irrelevant information. However, it poses a computational challenge: the convolution kernels become input-dependent, preventing parallel convolution operations. To overcome this problem, Mamba employs a hard-aware algorithm to speed up computation, leveraging three classical techniques, i.e., parallel associative scan (also called parallel prefix-sum)~\cite{harris2007parallel}, kernel fusion, and recomputation. Researchers observed that each state represents the whole compression of the previous states, enabling the direct computation of new states from previous ones. This suggests that the execution order of operations is independent of their associated attributes. Thus, Mamba employs the selective scan algorithm by computing sequences in segments, iteratively combining them, and integrating parameters that are conditioned on the input. In addition, GPUs consist of numerous processors capable of highly parallel computations. Mamba leverages the GPU's high-bandwidth memory (HBM) and fast SRAM I/O to avoid frequent SRAM writes to HBM through kernel fusion. Concretely, the discretization and recursive operations are performed in the higher-speed SRAM memory, and the output is then written back to the HBM. During backpropagation, the intermediate states are not stored but are recomputed when inputs are loaded from the HBM to the SRAM.

Although these modifications have enhanced model performance, granting it ``selective'' capabilities, they do not alter the inherent nature of SSMs, which operate in a unidirectional manner such as RNNs~\cite{rumelhart1986learning, hochreiter1997long}. This is not an issue in training large language models, as many of these models are trained in an autoregressive manner~\cite{radford2018improving,dai2019transformer}. However, for non-autoregressive speech models, we require a module with non-causal capabilities similar to attention. Thus, finding a suitable method to address this issue is essential. Moreover, the observation that Mamba's $\mathbf{A}$ matrix randomization and real-valued diagonal initialization perform equivalently suggests that Mamba's ability to delineate dependencies between inputs, compared to S4, needs further enhancement.

\subsection{Mamba for Audio Processing}
Recent works have explored the use of SSMs and Mamba methods in speech enhancement and separation. Distilled Audio SSM (DASS) applied knowledge distillation with SSMs and outperformed transformer-based models with processing sequences up to 2.5 hours long~\cite{bhati2024dass}. Audio Mamba (AuM)~\cite{erol2024audio} and its variant for audio tagging~\cite{lin2024audio} achieved comparable or better performance than Audio Spectrogram Transformers (AST) with roughly one-third of the parameters. Meanwhile, Self-Supervised Audio Mamba (SSAM)~\cite{yadav2024audio} established that unidirectional Mamba blocks were particularly effective for masked spectrogram modeling, consistently outperforming transformer-based approaches while better handling varying sequence lengths. \cite{semamba} proposed a SEMamba method for speech enhancement, which improved state-of-the-art performance by integrating time-frequency Mamba into advanced SE architecture. These advances collectively demonstrate the potential of Mamba architectures as an efficient and scalable alternative to transformers for audio processing tasks.

\cite{miyazaki2024exploring} conducted comprehensive experiments to compare Mamba encoder and decoder with existing methods across various tasks, including TTS, ASR, SLU, and SUMM. Mamba encoder and decoder are introduced. \cite{jiang2024speech} incorporates the Mamba module in state-of-the-art methods of speech separation, ASR, and TTS, comparing them with the vanilla models in their respective tasks. \cite{semamba} proposed a SEMamba method for speech enhancement, which improved state-of-the-art performance by integrating time-frequency Mamba into advanced SE architecture.
Our work, similar to \cite{miyazaki2024exploring} and \cite{jiang2024speech}, explores the use of Mamba in speech processing tasks. However, we specifically investigate Mamba's ability to learn speech information at various levels of abstraction, positioning it as an alternative to the self-attention module within Transformer and its derivatives. Beyond experiment-based analysis, we delve into the reasons behind the independent Mamba model's lower performance in ASR by examining the impact of nonlinearity.

\section{Investigating Mamba in Speech Processing} \label{rule}
\subsection{Bidirectional processing}
The original Mamba performs causal computations in a unidirectional manner, using only historical information. However, in speech tasks, the model is provided with the complete speech signal. Therefore, Mamba requires bidirectional computations, as employed in the~MHSA module, to capture global dependencies within the features of the input signal. In this paper, we explored two bidirectional strategies for Mamba in speech tasks, i.e., inner bidirectional Mamba~(InnBiMamba) from vision Mamba~\cite{zhu2024vision} and external bidirectional Mamba~(ExtBiMamba) as shown in Figure~\ref{Bi-directional}.

\textbf{InnBiMamba}. We first explore the inner bidirectional Mamba~(InnBiMamba) from Vision Mamba~\cite{zhu2024vision} for speech tasks as detailed in Figure~\ref{inner}. Here, two SSM modules share the same input and output projection layers. The process feeds the input forward into one SSM module, while reversing the input along the time dimension before feeding it into the other SSM module. The output of the backward SSM module is reversed back before being combined with the output of the forward SSM module. The combined output then passes through the output projection layer.

\textbf{ExtBiMamba}. We next propose a simpler and more straight bidirectional modeling strategy, i.e., external bidirectional Mamba (ExtBiMamba). The ExtBiMamba design is motivated by several key considerations in bidirectional sequence modeling. While existing approaches often rely on internal bidirectional mechanisms, our external approach offers distinct advantages. First, by maintaining separate input and output projections for forward and backward directions, ExtBiMamba allows each direction to learn direction-specific feature transformations that are optimal for processing the sequence in that particular order. This is particularly valuable for modeling contextual dependencies in speech signals, where certain acoustic patterns may be more interpretable when traversed in one direction rather than the other. 

The separation of forward and backward paths also enables more explicit modeling of temporal dependencies. The forward pathway captures the natural progression of acoustic events, while the backward pathway can effectively utilize future information when analyzing the sequence in reverse. This is critical for tasks that rely on contextual and sequential information, like ASR, speech enhancement, or emotion recognition. Compared to the InnBiMamba, the proposed ExtBiMamba design allows the forward and backward directions to learn complementary and non-redundant feature spaces. This separation enhances the capacity of BiMamba to represent complex, hierarchical structures in speech and contributes to improved performance in tasks requiring fine-grained temporal understanding. Detailed algorithms of these two methods are included in Section~\ref{Algorithms of Bidirectional Mamba}.

\subsection{Task-aware model designs}
Recent works have investigated Mamba in speech separation and speech enhancement. These tasks primarily focus on low-level spectral information of a speech signal~\cite{chen2022wavlm}. In contrast, other speech tasks like speech recognition and spoken language understanding require capturing high-level semantic information. With reference to equation~\ref{mamba_euqation}, SSM comprises mostly linear computations. This implies that it has limited capability to capture high-level information such as semantics and emotions. Although SiLU is used within residual structures in practical implementations, this is primarily to represent the parameter D in equation~\ref{continue_equation} for the state space model~\cite{gu2023mamba}. Therefore, adding more nonlinearity ability is crucial for Mamba to capture high-level information. 

To capture information of various abstraction levels, we progressively explored three structures for increasing nonlinearity capability. As depicted in Figure~\ref{Pure Mamba}, the first strategy uses the Mamba/BiMamba layers independently~(i.e., as a direct replacement for the transformer layer) to construct the Mamba/BiMamba model. The second approach employs the Mamba/BiMamba layer to replace the MHSA modules within the Transformer, where the feed-forward network (FFN) and layer normalization are used to provide nonlinearity. The third replaces the MHSA modules with Mamba/BiMamba layers in the Conformer, which is a variant of the Transformer employing a convolutional layer after each MHSA designed to additionally capture local information.

\subsection{Algorithms of Bidirectional Mamba modules}
\label{Algorithms of Bidirectional Mamba}

\begin{algorithm}
\caption{InnBiMamba Layer Workflow}
\label{alg:Inner bidirectional Mamba}
\begin{algorithmic}[1]
\REQUIRE $\mathbf{H}_{l-1}$: \textcolor{black}{$(B,L,D)$}
\ENSURE $\mathbf{H}_{l}$: \textcolor{black}{$(B,L,D)$}
    \STATE \textcolor{black}{$\mathbf{H}_{l-1}^{\prime}: \textcolor{black}{(B,L,D)} \gets \textbf{Norm}(\mathbf{H}_{l-1})$}
    \STATE \textcolor{violet}{$\mathbf{x}: \textcolor{black}{(B,L,E)} \gets \textbf{Linear}^{\mathbf{x}}(\mathbf{H}_{l-1}^{\prime})$}
    \STATE \textcolor{violet}{$\mathbf{z}: \textcolor{black}{(B,L,E)} \gets \textbf{Linear}^{\mathbf{z}}(\mathbf{H}_{l-1}^{\prime})$}
    
\FOR{$o \in \{\text{forward}, \text{backward}\}$}
    \STATE $\mathbf{x}_o^{\prime}: \textcolor{black}{(B,L,E)} \leftarrow\textbf{SiLU}(\textbf{Conv1d}_{o}(\mathbf{x}))$
    \STATE $\mathbf{B}_{o}: \textcolor{black}{(B,L,N)} \leftarrow \textbf{Linear}_{o}^{\mathbf{B}}(\mathbf{x}_{o}^{\prime})$
    \STATE $\mathbf{C}_{o}: \textcolor{black}{(B,L,N)} \leftarrow \textbf{Linear}_{o}^{\mathbf{C}}\left({\mathbf{x}_{o}}^{\prime}\right)$
    \STATE $\boldsymbol{\Delta}_o: \textcolor{black}{(B,L,E)} \leftarrow \log \left(1+\exp \left(\right.\right. \textbf{Linear}_{o}^{\Delta}\left(\mathbf{x}_o^{\prime}\right)+ \textbf{Parameter}_{o}^{\Delta}))$
    \STATE $\overline{\mathbf{A}}_{o}: \textcolor{black}{(B,L,E,N)} \leftarrow \boldsymbol{\Delta}_{o} \boldsymbol{\bigotimes} \textbf{ Parameter}_o^{\mathbf{A}}$
    \STATE $\overline{\mathbf{B}}_{o}: \textcolor{black}{(B,L,E,N)} \leftarrow \boldsymbol{\Delta}_{o} \boldsymbol{\bigotimes} \, \mathbf{  B}_{o}$
    \STATE  $\mathbf{y}_{o}: \textcolor{black}{(B,L,E)} \leftarrow \mathbf{S S M}\left(\overline{\mathbf{A}}_o, \overline{\mathbf{B}}_{o}, \mathbf{C}_{o}\right)\left(\mathbf{x}_{o}^{\prime}\right)$
    % \STATE \textcolor{violet}{$\mathbf{y}_{o}^{\prime}: \texttt{(B,L,E)} \leftarrow \textbf{Linear}_{o}^{\mathbf{T}}(\mathbf{y}_{o} \bigodot \textbf{SiLU}(\mathbf{z}_{o}))$}
\ENDFOR
\STATE \textcolor{violet}{$y^{\prime}_{\text{forward}}: \textcolor{black}{(B,L,E)} \gets y_{\text{forward}} \odot \textbf{SiLU}(\mathbf{z})$}
\STATE \textcolor{violet}{$y^{\prime}_{\text{backward}}: \textcolor{black}{(B,L,E)} \gets y_{\text{backward}} \odot \textbf{SiLU}(\mathbf{z})$}
\STATE \textcolor{violet}{$\mathbf{H}_{l} : \textcolor{black}{(B,L,D)} \gets \textbf{Linear}^{\mathbf{H}}(y^{\prime}_{\text{forward}} +y^{\prime}_{\text{backward}}) + \mathbf{H}_{l-1}$}
\RETURN $\mathbf{H}_{l}$
\end{algorithmic}
\end{algorithm}

\begin{algorithm}
\caption{ExtBiMamba Layer Workflow}
\label{alg:External bidirectional Mamba}
\begin{algorithmic}[1]
\REQUIRE $\mathbf{H}_{l-1}$: \textcolor{black}{$(B,L,D)$}
\ENSURE $\mathbf{H}_{l}$: \textcolor{black}{$(B,L,D)$}
\STATE \textcolor{black}{$\mathbf{H}_{l-1}^{\prime}: \textcolor{black}{(B,L,D)} \gets \textbf{Norm}(\mathbf{H}_{l-1})$}

\FOR{$o \in \{\text{forward}, \text{backward}\}$}
    \STATE \textcolor{violet}{$\mathbf{x}_{o}: \textcolor{black}{(B,L,E)} \gets \textbf{Linear}_{o}^{\mathbf{x}}(\mathbf{H}_{l-1}^{\prime})$}
    \STATE \textcolor{violet}{$\mathbf{z}_{o}: \textcolor{black}{(B,L,E)} \gets \textbf{Linear}_{o}^{\mathbf{z}}(\mathbf{H}_{l-1}^{\prime})$}
    \STATE $\mathbf{x}_o^{\prime}: \textcolor{black}{(B,L,E)} \leftarrow\textbf{SiLU}(\textbf{Conv1d}_{o}(\mathbf{x}_{o}))$
    \STATE $\mathbf{B}_{o}: \textcolor{black}{(B,L,N)} \leftarrow \textbf{Linear}_{o}^{\mathbf{B}}(\mathbf{x}_{o}^{\prime})$
    \STATE $\mathbf{C}_{o}: \textcolor{black}{(B,L,N)} \leftarrow \textbf{Linear}_{o}^{\mathbf{C}}\left({\mathbf{x}_{o}}^{\prime}\right)$
    \STATE $\boldsymbol{\Delta}_o: \textcolor{black}{(B,L,E)} \leftarrow \log \left(1+\exp \left(\right.\right. \textbf{Linear}_{o}^{\Delta}\left(\mathbf{x}_o^{\prime}\right)+ \textbf{Parameter}_{o}^{\Delta}))$
    \STATE $\overline{\mathbf{A}}_{o}: \textcolor{black}{(B,L,E,N)} \leftarrow \boldsymbol{\Delta}_{o} \boldsymbol{\bigotimes} \textbf{ Parameter}_o^{\mathbf{A}}$
    \STATE $\overline{\mathbf{B}}_{o}: \textcolor{black}{(B,L,E,N)} \leftarrow \boldsymbol{\Delta}_{o} \boldsymbol{\bigotimes} \, \mathbf{  B}_{o}$
    \STATE  $\mathbf{y}_{o}: \textcolor{black}{(B,L,E)} \leftarrow \mathbf{S S M}\left(\overline{\mathbf{A}}_o, \overline{\mathbf{B}}_{o}, \mathbf{C}_{o}\right)\left(\mathbf{x}_{o}^{\prime}\right)$
    \STATE \textcolor{violet}{$\mathbf{y}_{o}^{\prime}: \textcolor{black}{(B,L,D)} \leftarrow \textbf{Linear}_{o}^{\mathbf{H}}(\mathbf{y}_{o} \bigodot \textbf{SiLU}(\mathbf{z}_{o}))$}
\ENDFOR
\STATE \textcolor{violet}{$\mathbf{H}_{l}: \textcolor{black}{(B,L,D)}\gets (y^{\prime}_{\text{forward}} +y^{\prime}_{\text{backward}}) + \textbf{H}_{l-1}$}
\RETURN $\mathbf{H}_{l}$
\end{algorithmic}
\end{algorithm}

\textcolor{black}{Algorithms~\ref{alg:Inner bidirectional Mamba} and~\ref{alg:External bidirectional Mamba} illustrate the workflows of InnBiMamba and ExtBiMamba, respectively. The main differences are highlighted in violet. For the InnBiMamba layer, the linear input projections ($\textbf{Linear}^{\mathbf{x}}$ and $\textbf{Linear}^{\mathbf{z}}$) and the linear output projection $\textbf{Linear}^{\mathbf{H}}$ are shared across forward and backward operations. In contrast, the ExtBiMamba layer uses different input linear projections ($\textbf{Linear}_{o}^{\mathbf{x}}$ and $\textbf{Linear}_{o}^{\mathbf{z}}$) and output linear projections ($\textbf{Linear}_{o}^{\mathbf{H}}$) across forward and backward operations. \textcolor{black}{The dimensionality of the matrices is defined as follows: $B$ represents the training batch size, $L$ denotes the sequence length, \textcolor{black}{$D$ denotes the dimension of the input hidden state, and $E$ denotes the expanded hidden state dimension,} and $N$ denotes the SSM state dimension.}
% The input sequence $\mathbf{H}_{l-1}$ is first processed through a normalization operation. InnBiMamba shares the same linear input layers across forward and backward directions, which transform the normalized input $\mathbf{H}_{l-1}^{\prime}$ to $\mathbf{x}$ and $\mathbf{z}$. For each direction, the $\mathbf{x}$ is firstly fed into a 1-D convolution layer followed by the SiLU activation to produce $\mathbf{x}_{o}^{\prime}$. The $x_{0}^{\prime}$ is then linearly transformed to $\mathbf{B}_{o}$
}

\section{Experimental Setup}
\subsection{Speech Enhancement}

\textbf{Datasets}. We follow previous works~\cite{zhangtfa,zhang2024} and employ the clean speech data from the LibriSpeech~\textit{train-clean-100} corpus~\cite{panayotov2015librispeech} as the training set, containing $28\,539$ speech clips spoken by $251$ speakers. The noise recordings are collected from the following datasets~\cite{tfaj}, i.e., the noise data of the MUSAN datasets~\cite{snyder2015musan}, the RSG-10 dataset~\cite{steeneken1988description} (\textit{voice babble}, \textit{F16}, and \textit{factory welding} are excluded for testing), the Environmental Noise dataset~\cite{saki2016automatic,saki2016smartphone}, the colored noise set (with an $\alpha$ value ranging from -2 to 2 in increments of 0.25)~\cite{deepmmse}, the UrbanSound dataset~\cite{Urban} (\textit{street music} recording no 26 270 is excluded for testing), the QUT-NOISE dataset~\cite{dean2010qut}, and the Nonspeech dataset~\cite{hu2010tandem}.

Noise recordings that exceeds $30$ seconds in duration are split into clips of $30$ seconds or less. This yields $6\,809$ noise clips, with each clip less than or equal to $30$ seconds in duration. For validation experiments, $1\,000$ clean speech and noise clips (without replacement) were randomly drawn from the aforementioned clean speech and noise sets and mixed to generate a validation set of $1\,000$ noisy clips, where each clean speech clip was degraded by a random section of one noise clip at a random SNR level (sampled between $-10$ and $20$ dB, in 1 dB steps). For evaluation experiments, we employed four real-world noise sources (excluded from the training set) including two non-stationary and two colored ones. The two non-stationary noise sources were the \textit{voice babble} from the RSG-10 noise dataset~\cite{steeneken1988description} and \textit{street music} from the Urban Sound dataset~\cite{Urban}. The two colored noise sources were \textit{F16} and \textit{factory welding} from RSG-10 noise dataset~\cite{steeneken1988description}. For each of the four noises, we randomly picked twenty clean speech clips (without replacement) from the \textit{test-clean-100} of LibriSpeech corpus~\cite{panayotov2015librispeech} and degraded each clip with a random section of the noise clip at the five SNR levels, i.e., $\left\{ -5 \text{ dB}, 0 \text{ dB}, 5 \text{ dB}, 10 \text{ dB}, 15 \text{ dB} \right\}$. This generated $400$ noisy mixtures for evaluation.

\textbf{Feature Extraction}. All audio signals are sampled at a rate of 16 kHz. We employ a 512-sample (32 ms) long square-root-Hann window with a hop length of 256 samples (16 ms), to extract a 257-point single-sided STFT spectral magnitude as the input to the neural models~\cite{tfaj}.

\textbf{Model Configurations}. In our experiments, we employ the same backbone network architecture (a typical neural solution to speech enhancement)~\cite{tfaj,zhang2024,ripple,mhanet}, which comprises an input embedding layer, stacked feature transformation layers (such as Mamba, Transformer, and Conformer layers), and an output layer. To systematically study the Mamba networks, we use the standard Transformer~\cite{tfaj,zhang2024} and Conformer~\cite{gulati20_interspeech} models as the baseline backbone networks, across causal and non-causal configurations. 

\begin{table}[!t]
\centering
\scriptsize % Further reduce the font size
\setlength{\tabcolsep}{2.6pt} % Adjust column separation for tighter fit
\def\arraystretch{1.1} % Slightly reduce row height
\setlength{\abovetopsep}{0pt}
\setlength\belowbottomsep{0pt} 
\setlength\aboverulesep{0pt} 
\setlength\belowrulesep{0pt}

\caption{Implementation details in different tasks and datasets. We show the best Transformer configurations reported in ESPnet Official}
\begin{tabular}{lcccc}
\toprule[1.0pt]
& \textbf{LibriSpeech100} & \textbf{LibriSpeech960} & \textbf{ASRU} & \textbf{SEAME} \\
\midrule
\multicolumn{5}{l}{\textbf{Frontend}} \\
window length & 400 & 400 & 400 & 400 \\
hop length & 160 & 128 & 160 & 160 \\
\midrule
\multicolumn{5}{l}{\textbf{SpecAug}} \\
time warp window & 5 & 5 & 5 & 5 \\
num of freq masks & 2 & 2 & 2 & 2 \\
freq mask width & (0, 27) & (0, 30) & (0, 27) & (0, 27) \\
num of time masks & 2 & 2 & 2 & 2 \\
time mask width & (0, 0.05) & (0, 40) & (0, 0.05) & (0, 0.05) \\
\midrule
\multicolumn{5}{l}{\textbf{Architecture}} \\
feature size $d$ & 256 & 512 & 256 & 256 \\
hidden size $d_{\text{hidden}}$ & 1024 & 2048 & 1024 & 2048 \\
attention heads $h$ & 4 & 8 & 4 & 4 \\
num of encoder layers & 18 & 18 & 18 & 24 \\
depth-wise conv kernel & 31 & 31 & 31 & 31 \\
\midrule
\multicolumn{5}{l}{\textbf{Training}} \\
epochs & 70 & 100 & 70 & 70 \\
learning rate & 2e-3 & 2e-3 & 2e-3 & 2e-3 \\
warmup steps & 15k & 25k & 15k & 15k \\
weight decay & 1e-6 & 1e-6 & 1e-6 & 1e-6 \\
dropout rate & 0.1 & 0.1 & 0.1 & 0.1 \\
ctc weight & 0.3 & 0.3 & 0.3 & 0.3 \\
label smoothing & 0.1 & 0.1 & 0.1 & 0.1 \\
\bottomrule[1.0pt]
\end{tabular}
% \vspace{-1.0em}
\label{Transformer config for ASR}
\end{table}

\begin{table}[!t]
\centering
\scriptsize % Reduce the font size
\setlength{\tabcolsep}{2.6pt} % Adjust column separation for a tighter fit
\def\arraystretch{1.1} % Reduce row height
\setlength{\abovetopsep}{0pt}
\setlength\belowbottomsep{0pt} 
\setlength\aboverulesep{0pt} 
\setlength\belowrulesep{0pt}

\caption{Implementation details in different tasks and datasets. We show the best Conformer configurations reported in ESPnet Official }
\begin{tabular}{lcccc}
\toprule[1.0pt]
& \textbf{LibriSpeech100} & \textbf{LibriSpeech960} & \textbf{ASRU} & \textbf{SEAME} \\
\midrule
\multicolumn{5}{l}{\textbf{Frontend}} \\
window length & 400 & 400 & 400 & 400 \\
hop length & 160 & 160 & 160 & 160 \\
\midrule
\multicolumn{5}{l}{\textbf{SpecAug}} \\
time warp window & 5 & 5 & 5 & 5 \\
num of freq masks & 2 & 2 & 2 & 2 \\
freq mask width & (0, 27) & (0, 27) & (0, 30) & (0, 30) \\
num of time masks & 2 & 2 & 2 & 2 \\
time mask width & (0, 0.05) & (0, 0.05) & (0, 40) & (0, 40) \\
\midrule
\multicolumn{5}{l}{\textbf{Architecture}} \\
feature size $d$ & 256 & 512 & 256 & 256 \\
hidden size $d_{\text{hidden}}$ & 1024 & 2048 & 2048 & 2048 \\
attention heads $h$ & 4 & 8 & 4 & 4 \\
num of encoder layers & 12 & 12 & 12 & 12 \\
depth-wise conv kernel & 31 & 31 & 31 & 31 \\
\midrule
\multicolumn{5}{l}{\textbf{Training}} \\
epochs & 120 & 50 & 70 & 70 \\
learning rate & 2e-3 & 2.5e-3 & 1e-3 & 1e-3 \\
warmup steps & 15k & 40k & 25k & 25k \\
weight decay & 1e-6 & 1e-6 & 1e-6 & 1e-6 \\
dropout rate & 0.1 & 0.1 & 0.1 & 0.1 \\
ctc weight & 0.3 & 0.3 & 0.3 & 0.3 \\
label smoothing & 0.1 & 0.1 & 0.1 & 0.1 \\
\bottomrule[1.0pt]
\end{tabular}
\label{Conformer config for ASR}
\vspace{-1.0em}
\end{table}

\begin{table}[t]
\centering
\scriptsize % Reduce the font size
\setlength{\tabcolsep}{2pt} % Adjust column separation for a tighter fit
\def\arraystretch{1.1} % Slightly reduce row height
\setlength{\abovetopsep}{0pt}
\setlength\belowbottomsep{0pt} 
\setlength\aboverulesep{0pt} 
\setlength\belowrulesep{0pt}
\caption{Implementation details in different tasks and datasets. We show the best branchformer configurations reported in ESPnet Official }
\begin{tabular}{lcccc}
\toprule[1.0pt]
& \textbf{LibriSpeech100} & \textbf{LibriSpeech960} & \textbf{ASRU} & \textbf{SEAME} \\
\midrule
\multicolumn{5}{l}{\textbf{Frontend}} \\
window length & 400 & 400 & 400 & 400 \\
hop length & 160 & 160 & 160 & 160 \\
\midrule
\multicolumn{5}{l}{\textbf{SpecAug}} \\
time warp window & 5 & 5 & 5 & 5 \\
num of freq masks & 2 & 2 & 2 & 2 \\
freq mask width & (0, 27) & (0, 27) & (0, 27) & (0, 27) \\
num of time masks & 2 & 2 & 2 & 2 \\
time mask width & (0, 0.05) & (0, 0.05) & (0, 0.05) & (0, 0.05) \\
\midrule
\multicolumn{5}{l}{\textbf{Architecture}} \\
feature size $d$ & 256 & 512 & 256 & 256 \\
hidden size $d_{\text{hidden}}$ & 1024 & 2048 & 2048 & 1024 \\
attention heads $h$ & 4 & 8 & 4 & 4 \\
num of encoder layers & 12 & 18 & 12 & 12 \\
depth-wise conv kernel & 31 & 31 & 31 & 31 \\
\midrule
\multicolumn{5}{l}{\textbf{Training}} \\
epochs & 70 & 70 & 70 & 70 \\
learning rate & 2e-3 & 2.5e-3 & 2e-3 & 2e-3 \\
warmup steps & 15k & 40k & 15k & 15k \\
weight decay & 1e-6 & 1e-6 & 1e-6 & 1e-6 \\
dropout rate & 0.1 & 0.1 & 0.1 & 0.1 \\
ctc weight & 0.3 & 0.3 & 0.3 & 0.3 \\
label smoothing & 0.1 & 0.1 & 0.1 & 0.1 \\
\bottomrule[1.0pt]
\end{tabular}
\label{Branchformer config for ASR}
\vspace{-1.0em}
\end{table}

To perform extensive comparison studies across different model sizes, we use the Transformer and Conformer architectures comprising $4$ and $6$ stacked Transformer and Conformer layers respectively~\cite{tfaj}. For the Transformer speech enhancement backbone, we follow the configuration in~\cite{zhang2024,tfaj}: the layer dimension $d_{\text{model}}\!=\!256$, the number of attention heads $H=8$, and the inner-layer size of the feed-forward network (FFN) $d_{f\!f}=1024$. For the Conformer backbone, we adopt the parameter configurations in~\cite{tfaj}: the layer dimension $d_{\text{model}}\!=\!256$, the number of attention heads $H=8$, the kernel size of convolution $32$, the expansion factor for convolution module $2$, and the inner-layer size of FFN $d_{f\!f}=1024$. For Mamba and BiMamba models, we employ the following hyper-parameter configurations: the hidden state dimension $D=256$, the SSM state dimension $N\!=\!16$, the local convolution width $d_{\text{conv}}\!=\!4$, and the expanded hidden state dimension $E=512$ (i.e., the expansion factor $E_{f}=2$). All the experiments were run on an NVIDIA Tesla V100-SXM2-32GB GPU.

\textbf{Training strategy}. We use mean-square error (MSE) on the power-law compressed spectral magnitude as the objective loss function~\cite{ephrat2018looking}. The noisy mixtures are dynamically generated at training time. For each mini-batch, we randomly pick 10 clean speech clips and degraded each clip by a random section of a random noise clip at a random SNR level sampled from $-10$ to $20$ dB (in $1$ dB steps). The \textit{Adam} optimization algorithm~\cite{Adam} is employed for gradient descent, with parameters as in~\cite{vaswani2017attention}, i.e., $\beta_{1}=0.9$, $\beta_{2}=0.98$, and $\epsilon=10^{-9}$. We utilize the gradient clipping technology to cut the gradient values to a range between $-1$ and $1$. All the models are trained for 150 epochs for fair comparison. The warm-up strategy is adopted to adjust the learning rate: $lr = d_{model}^{-0.5}\cdot \textrm{min} \left(n\_step^{-0.5}, n\_step \cdot w\_steps^{-1.5}\right)$, where $n\_step$ and $w\_steps$ denote the iteration steps and the warm-up iteration steps, respectively. We follow the study~\cite{tfaj} and set $w\_steps$ as $40000$.

\textbf{Evaluation Metrics}. We evaluate enhanced speech with five commonly used assessment metrics, i.e, perceptual evaluation of speech quality (PESQ)~\cite{pesq}, extended short-time objective intelligibility (ESTOI), and three composite metrics. For PESQ, both wide-band PESQ (W-PESQ) and narrow-band PESQ (N-PESQ) were used to evaluate the speech quality, with the score range of $[-0.5,\,4.5]$. The ESTOI~\cite{jensen2016algorithm} score is typically between 0 and 1. The three composite metrics~\cite{hu2007evaluation} are used to predict the mean opinion scores of the intrusiveness of background noise (CBAK), the signal distortion (CSIG), and the overall signal quality (COVL), respectively, with the score range of $[0,\,5]$. For each metric, $\uparrow$ indicates that higher values are preferable, while $\downarrow$ indicates that lower values are preferable.

\subsection{Speech Recognition}
\textbf{Datasets}. We evaluate our models on ASR with four datasets, i.e., LibriSpeech~\cite{panayotov2015librispeech}, AN4~\cite{acero1990environmental}, SEAME~\cite{lyu2010seame}, and ASRU~\cite{shi2020asru}, in which all speech signals are sampled at 16 kHz. LibriSpeech (LibriSpeech960) containing approximately 1000 hours of audio recordings and their paired texts, in which a subset LibriSpeech100 is used for ablation studies due to its higher recording quality. The AN4 dataset contains approximately one hour of audio recordings of primarily spoken alphanumeric strings, such as postal codes and telephone numbers. It is employed to assess the model’s ability to perform with a small dataset. Two English-Mandarin code-switching datasets SEAME and ASRU-CS-2019 (denoted as ASRU) are then used for a more challenging scenario compared to monolingual. The SEAME dataset contains 200-hour spontaneous South-east Asian-accented speech with intra- and inter-sentential code-switches, divided as introduced in~\cite{zeng19_interspeech}. The ASRU dataset contains a 500-hour Mandarin and a 200-hour code-switching training sets recorded in mainland China, where only the code-switching set is used for training, following~\cite{lyu2010seame,shi2020asru,liu2024aligning}.

For Mamba and BiMamba models, we employ the default hyper-parameters~\cite{gu2023mamba}: the SSM state dimension $N\!=\!16$, the local convolution width $d_{\text{conv}}\!=\!4$, and the expansion factor $E_{f}\!=\!2$. To keep the same batch size as in the official implementation of ESPnet, the experiments on the LibriSpeech-960 dataset were performed on an NVIDIA A100 80GB. All the other experiments were run on an NVIDIA V100 32GB. The detailed model configurations for each dataset are given in Tables~\ref{Transformer config for ASR},~\ref{Conformer config for ASR} and~\ref{Branchformer config for ASR}.

\textbf{Evaluation Metrics}. We employ word error rate~(WER) and mixed word error rate~(MER) to measure the ASR performance for monolingual and code-switching ASR tasks, respectively, where the MER considers the word error rate for English and the character error rate for Mandarin. We employ $\downarrow$ to indicate that lower WER and MER are preferable. All experiments are performed using the ESPnet toolkit~\cite{watanabe2018espnet}

\section{Experimental Results and Analysis} \label{Main Result}

\subsection{Speech Enhancement}
\textbf{InnBiMamba vs. ExtBiMamba}. Table~\ref{tab:bimamba} presents the comparison results of InnBiMamba and ExtBiMamba across different model sizes, in terms of six metrics, i.e., N-PESQ, W-PESQ, ESTOI, CSIG, CBAK, and COVL. Overall, ExtBiMamba consistently demonstrates slightly superior performance compared to InnBiMamba across the model sizes. In addition, Table~\ref{tab:bimamatime} presents the comparison results of InnBiMamba and ExtBiMamba in training time (time per training step, sec/step), inference speed, and computational complexity. The inference speed and computation complexity are measured in terms of real-time factor (RTF)~\cite{cleanunet,tfaj} and the number of multiply-accumulate operations (MACs), respectively. RTF is computed as the ratio of processing time to speech duration, measured on an NVIDIA Tesla V100 GPU and averaged over 20 runs. We employ a batch size of 4 speech utterances, each with a duration of 10 seconds~\cite{cleanunet}. The results demonstrate that ExtBiMamba achieves faster training and inference speeds and lower computational complexity compared to InnBiMamba at similar model sizes.

\begin{table}[!htbp]
    \centering
    \scriptsize
    \def\arraystretch{1.3}
    \setlength{\tabcolsep}{1pt}
    \setlength{\abovetopsep}{0pt}
    \setlength\belowbottomsep{0pt} 
    \setlength\aboverulesep{0pt} 
    \setlength\belowrulesep{0pt}
    \caption{\textcolor{black}{The comparison results of InnBiMamba and ExtBiMamba network architectures in N-PESQ, W-PESQ, ESTOI (\%), CSIG, CBAK, and COVL. The model name is denoted in the format `network architecture–number of building blocks'.}}
    \label{tab:bimamba}
    \resizebox{\columnwidth}{!}{
    \begin{tabular}{@{}lcccccccc@{}}
        \toprule[1.0pt]
        \multirow{2}{*}{\textcolor{black}{\textbf{Model}}} & \multirow{2}{*}{\textbf{\#Para.}} & \multirow{2}{*}{\textbf{Cau.}} & \multicolumn{6}{c}{\textbf{Metrics}}\\  
        \cline{4-9}
        & & & \textcolor{black}{N-PESQ$\uparrow$} & \textcolor{black}{W-PESQ$\uparrow$} & \textcolor{black}{ESTOI$\uparrow$ }& \textcolor{black}{CSIG $\uparrow$} & \textcolor{black}{CBAK$\uparrow$} & \textcolor{black}{COVL$\uparrow$} \\  
        \midrule
        Noisy  & -- & -- & \textcolor{black}{1.88} & \textcolor{black}{1.24} & \textcolor{black}{56.12} & \textcolor{black}{2.26} & \textcolor{black}{1.80} & \textcolor{black}{1.67} \\
        InnBiMamba-9  &  4.48M  & \ding{55} & 2.84 & 2.14 & 75.74 & 3.39 & 2.59 & 2.74  \\
        ExtBiMamba-5  &  4.51M  & \ding{55} & \bf 2.86 & \bf 2.15 & \bf 76.12 & \bf 3.46 & \bf 2.60 & \bf 2.78  \\
        \hdashline
        InnBiMamba-13  &  6.41M  & \ding{55} & \bf 2.90 & 2.19 & 76.89 & 3.50 & 2.63 & 2.82  \\
        ExtBiMamba-7  &  6.26M  & \ding{55} & \bf 2.90 & \bf 2.20 & \bf 77.04 & \bf 3.54 & \bf 2.64 & \bf 2.84  \\
        \toprule[1.0pt]
    \end{tabular}}
    \vspace{-0.7em}
\end{table}

\begin{table}[!htbp]
    \centering
    % \scriptsize
    % \footnotesize
    \small
    \def\arraystretch{1.25}
    \setlength{\tabcolsep}{7.0pt} % Reduce column padding
    \setlength{\abovetopsep}{0pt}
    \setlength\belowbottomsep{0pt} 
    \setlength\aboverulesep{0pt} 
    \setlength\belowrulesep{0pt}
    \caption{\textcolor{black}{The comparison results of InnBiMamba and ExtBiMamba in terms of training speed measured by time per training step (seconds/step), inference speed measured by RTF ($\times 10^{-4}$), and computation complexity (MACs).}}
    \label{tab:bimamatime}
    % \begin{tabular}{@{}lcc||cc@{}}
    % \toprule[1.0pt]
    %  & \textcolor{black}{InnBiMamba-9} & \textcolor{black}{ExtBiMamba-5}  & \textcolor{black}{InnBiMamba-13} & \textcolor{black}{ExtBiMamba-7}\\
    %  \hline
    % sec/step & 0.159 &  0.122 & 0.212 &  0.159 \\
    % % RTF & $1.88\times 10^{-4}$ & $1.69\times 10^{-4}$ &  $2.76\times 10^{-4}$  &  $2.34\times 10^{-4}$ \\
    % RTF & $1.88$ & $1.69$ &  $2.76$  &  $2.34$ \\
    % MACs & 3.09G & 2.98G & 4.43G  & 4.14G \\
    % \toprule[1.0pt]
    % \end{tabular}
    % \vspace{-1.0em}

    \begin{tabular}{@{}lcccc@{}}
    \toprule[1.0pt]
    \textcolor{black}{\textbf{Model}} & \textbf{\#Para.} & \textcolor{black}{sec/step$\downarrow$} & \textcolor{black}{RTF$\downarrow$} & \textcolor{black}{MACs$\downarrow$} \\
    \hline
    \textcolor{black}{InnBiMamba-9}  & 4.48M & 0.159 & 1.88 & \textcolor{black}{3.09G} \\
    \textcolor{black}{ExtBiMamba-5}  & 4.51M & \textbf{0.122} & \textbf{1.71 }& \textcolor{black}{\textbf{2.98G}} \\
    \hdashline
    \textcolor{black}{InnBiMamba-13} & 6.41M & 0.212 & 2.76 & \textcolor{black}{4.43G} \\
    \textcolor{black}{ExtBiMamba-7}  & 6.26M & \textbf{0.159} & \textbf{2.34 }& \textcolor{black}{\textbf{4.14G}} \\
    \toprule[1.0pt]
    \end{tabular}
    \vspace{-1.0em}
\end{table}

% InnBiMamba-9              3.091G/6.182G/12.365G
% InnBiMamba-13             4.428G/8.857G/17.713G
% BiMamba-5                 2.979G/5.958G/11.916G
% BiMamba-6                 3.558G/7.116G/14.233G
% BiMamba-7                 4.137G/8.275G/16.550G
\begin{table}[t]
    \centering
    \scriptsize
    \def\arraystretch{1.3}
    \setlength{\tabcolsep}{1pt}
    \setlength{\abovetopsep}{0pt}
    \setlength\belowbottomsep{0pt} 
    \setlength\aboverulesep{0pt} 
    \setlength\belowrulesep{0pt}
    \caption{\textcolor{black}{Performance comparisons of Mamba to Transformer and Conformer network architectures in \textcolor{black}{N-PESQ, W-PESQ}, ESTOI (\%), CSIG, CBAK, and COVL, across causal and non-causal configurations. The model name is denoted in the format `network architecture–number of building blocks'.}}
    \label{tab:transformer_mamba}
    \resizebox{\columnwidth}{!}{
    \begin{tabular}{@{}lcccccccc@{}}
        \toprule[1.0pt]
        \multirow{2}{*}{\textbf{Model}} & \multirow{2}{*}{\textbf{\#Para.}} & \multirow{2}{*}{\textbf{Cau.}} & \multicolumn{6}{c}{\textbf{Metrics}}\\  
        \cline{4-9}
        & & & \textcolor{black}{N-PESQ$\uparrow$} & \textcolor{black}{W-PESQ$\uparrow$} & \textcolor{black}{ESTOI$\uparrow$} & \textcolor{black}{CSIG$\uparrow$} & \textcolor{black}{CBAK$\uparrow$} & \textcolor{black}{COVL$\uparrow$} \\  
        \midrule
        Noisy  & -- & -- & \textcolor{black}{1.88} & \textcolor{black}{1.24} & \textcolor{black}{56.12} & \textcolor{black}{2.26} & \textcolor{black}{1.80} & \textcolor{black}{1.67} \\
        \hline
        \multicolumn{9}{l}{\textcolor{black}{\textbf{Mamba vs. Transformer}}} \\
        \hline
        Transformer-4 &  3.29M  & \ding{52}   & \textcolor{black}{2.56} & \textcolor{black}{1.84} & \textcolor{black}{70.32} & \textcolor{black}{3.17} & \textcolor{black}{2.39} & \textcolor{black}{2.47} \\
        Mamba-4       &  1.88M  & \ding{52}   & \textcolor{black}{2.60} & \textcolor{black}{1.87} & \textcolor{black}{70.99} & \textcolor{black}{3.17} & \textcolor{black}{2.41} & \textcolor{black}{2.48} \\
        Mamba-7       &  3.20M  & \ding{52}   & \textcolor{black}{\bf 2.64} & \textcolor{black}{\bf 1.91} & \textcolor{black}{\bf 72.46} & \textcolor{black}{\bf 3.26} & \textcolor{black}{\bf 2.45} & \textcolor{black}{\bf 2.56} \\
        \hdashline
        Transformer-4 & 3.29M  & \ding{55} & \textcolor{black}{2.74} & \textcolor{black}{2.01} & \textcolor{black}{73.44} & \textcolor{black}{3.31} & \textcolor{black}{2.50} & \textcolor{black}{2.63}  \\
        ExtBiMamba-3  &  2.76M  & \ding{55} & \textcolor{black}{2.76} & \textcolor{black}{2.05} & \textcolor{black}{73.93} & \textcolor{black}{3.36} & \textcolor{black}{2.53} & \textcolor{black}{2.67}  \\
        ExtBiMamba-4  &  3.64M  & \ding{55} & \bf 2.83 & \bf 2.11 & \bf 75.43 & \bf 3.46 & \bf 2.57 & \bf 2.75  \\
        \hdashline
        Transformer-6 & 4.86M  & \ding{55} & \textcolor{black}{2.78} & \textcolor{black}{2.05} & \textcolor{black}{74.56} & \textcolor{black}{3.38} & \textcolor{black}{2.52} & \textcolor{black}{2.69} \\
        ExtBiMamba-5  &  4.51M  & \ding{55} & 2.86 & 2.15 & 76.12 & 3.46 & 2.60 & 2.78  \\
        ExtBiMamba-6  &  5.39M  & \ding{55} & \bf 2.88 & \bf 2.17 & \bf 76.69 & \bf 3.50 & \bf 2.62 & \bf 2.82  \\
        \hline
        \multicolumn{9}{l}{\textcolor{black}{\textbf{Mamba vs. Conformer}}} \\
        \hline
        % \toprule[0.7pt]
        % \toprule[0.7pt]
        Conformer-4 &  6.22M  &  \ding{52}   & \textcolor{black}{2.67} & \textcolor{black}{1.94} & \textcolor{black}{72.78} & \textcolor{black}{3.30} & \textcolor{black}{2.46} & \textcolor{black}{2.59}  \\
        Mamba-13   &  5.83M  &  \ding{52}   & \textcolor{black}{\bf 2.70} & \textcolor{black}{\bf 1.97} & \textcolor{black}{\bf 73.53} & \textcolor{black}{\bf 3.32} & \textcolor{black}{\bf 2.50} & \textcolor{black}{\bf 2.62}  \\
        \hdashline
        Conformer-4 & 6.22M  & \ding{55} & \textcolor{black}{2.88} & \textcolor{black}{2.17} & \textcolor{black}{76.68} & \textcolor{black}{3.51} & \textcolor{black}{2.61} & \textcolor{black}{2.82}  \\
        ExtBiMamba-6  &  5.39M  & \ding{55} & 2.88 & 2.17 & 76.69 & 3.50 & 2.62 & 2.82  \\
        ExtBiMamba-7  &  6.26M  & \ding{55} & \bf 2.90 & \bf 2.20 & \bf 77.04 & \bf 3.54 & \bf 2.64 & \bf 2.84  \\
        \hdashline
        % \hline
        % \hline
        Conformer-6 &  9.26M  &  \ding{52}   & \textcolor{black}{2.68} & \textcolor{black}{1.94} & \textcolor{black}{73.41} & \textcolor{black}{3.30} & \textcolor{black}{2.46} & \textcolor{black}{2.59}  \\
        Mamba-20   &  8.89M  &  \ding{52}   & \textcolor{black}{\bf 2.72} & \textcolor{black}{\bf 2.00} & \textcolor{black}{\bf 74.09} & \textcolor{black}{\bf 3.35} & \textcolor{black}{\bf 2.51} & \textcolor{black}{\bf 2.65}  \\
        \hdashline
        Conformer-6 & 9.26M  & \ding{55} & \textcolor{black}{\bf 2.91} & \textcolor{black}{\bf 2.20} & \textcolor{black}{ 77.56} & \textcolor{black}{3.54} & \textcolor{black}{2.62} & \textcolor{black}{2.84}  \\
        ExtBiMamba-10  & 8.89M  & \ding{55} & \bf 2.91 & \bf 2.20 & \bf 77.64 & \bf 3.59 & \bf 2.65 & \bf 2.87 \\
        \toprule[1.0pt]
    \end{tabular}}
    \vspace{-1.5em}
\end{table}

\textbf{Mamba vs. Transformer}. Table~\ref{tab:transformer_mamba} reports the comparison results of Mamba with Transformer and Conformer network architectures. For Mamba models, we report the results of Mamba models with the same number of layers and a similar model size to the Transformer. It can be observed that the Mamba models consistently demonstrate obvious performance superiority over the Transformer models with lower parameter overheads, across causal and non-causal configurations. For instance, the Mamba-7 (3.20M) and ExtBiMamba-5 (4.51M) improve on the causal Transformer-4 (3.29M) and the non-causal Transformer-6 (4.81M) by 0.08 and 0.08, 0.07 and 0.1, 2.14\% and 1.56\%, 0.09 and 0.08, 0.06 and 0.08, and 0.09 and 0.09 in terms of N-PESQ, W-PESQ, ESTOI, CSIG, CBAK, and COVL, respectively. In addition, ExtBiMamba models provide substantial performance improvements over original (unidirectional) Mamba models across all the metrics, which confirms the effectiveness of the bidirectional modeling. ExtBiMamba-5 (4.51M) provided gains of 0.18 in N-PESQ, 0.21 in W-PESQ, 2.92\% in ESTOI, 0.16 in CSIG, 0.13 in CBAK, and 0.19 in COVL over Mamba-10 (4.51M), respectively. 

\textbf{Mamba vs. Conformer}. From the comparison results of Conformer and Mamba in Table~\ref{tab:transformer_mamba}, We can see that original Mamba (causal) models outperform causal Conformer models across all six metrics while involving fewer parameters. For instance, compared to causal Conformer-6 (9.26M), Mamba-20 (8.89M) improves N-PESQ by 0.04, W-PESQ by 0.06, ESTOI by 0.68\%, CSIG by 0.05, CBAK by 0.05, and COVL by 0.06. It can also be seen that overall, ExtBiMamba models exhibit slightly better or comparable performance to non-causal Conformer models. The substantial performance superiority of the bidirectional modeling is observed from evaluation results of ExtBiMamba-6 (5.39M) vs. Mamba-13 (5.83M) and ExtBiMamba-10 (8.89M) vs. Mamba-20 (8.89M).

\begin{table}[!t]
    \centering
    \scriptsize
    % \small
    % \def\arraystretch{1.05}
    \def\arraystretch{1.3}
    \setlength{\tabcolsep}{1.5pt}
    \setlength{\abovetopsep}{0pt}
    \setlength\belowbottomsep{0pt} 
    \setlength\aboverulesep{0pt} 
    \setlength\belowrulesep{0pt}
    % \vspace*{-0.1in}
    % \footnotesize	
    % \caption{Average CSIG scores for each SNR level and the highest CSIG scores are highlighted in boldface.}
    \caption{\textcolor{black}{The comparison results of training time (sec/step), RTF ($\times 10^{-4}$), and computation complexity (MACs), across different input speech lengths (10s, 20s, and 40s).}}
    \label{tab:complexity}
    % \vspace*{-0.1in}
    \resizebox{0.99\columnwidth}{!}{
    \begin{tabular}{@{}lccccccccc@{}}
    \toprule[1.0pt]
        
% Transformer-4             2.842G/7.275G/20.908G
% Transformer-6             4.222G/10.829G/31.197G
% Conformer-4               4.669G/10.927G/28.214G
% Conformer-6               6.962G/16.307G/42.156G
% BiMamba-3                 1.820G/3.641G/7.282G
% BiMamba-4                 2.400G/4.799G/9.599G
% BiMamba-5                 2.979G/5.958G/11.916G
% BiMamba-6                 3.558G/7.116G/14.233G
% BiMamba-7                 4.137G/8.275G/16.550G
% BiMamba-8                 4.717G/9.433G/18.867G
% BiMamba-10                5.875G/11.750G/23.501G
% ConExtBiMamba-6           8.26G/16.53G/33.06G
% ConInnBiMamba-6           6.795G/13.591G/27.182G
% TransExtBiMamba-6         5.526G/11.051G/22.102G
% TransInnBiMamba-6         4.056G/8.112G/16.223G
% InnBiMamba-9              3.091G/6.182G/12.365G
% InnBiMamba-13             4.428G/8.857G/17.713G
        \multirow{2}{*}{\textbf{Model}} & \multirow{2}{*}{\textbf{\#Para.}} & \multirow{2}{*}{\textcolor{black}{sec/step$\downarrow$}} & \multicolumn{2}{c}{\textcolor{black}{10s}} & \multicolumn{2}{c}{\textcolor{black}{20s}} & \multicolumn{2}{c}{\textcolor{black}{40s}} \\
        & & & \textcolor{black}{RTF$\downarrow$}& \textcolor{black}{MACs$\downarrow$} & \textcolor{black}{RTF$\downarrow$}& \textcolor{black}{MACs$\downarrow$} & \textcolor{black}{RTF$\downarrow$}& \textcolor{black}{MACs$\downarrow$} \\  
        \hline
        \multicolumn{7}{l}{\textcolor{black}{\textbf{Mamba vs. Transformer}}} \\
        \hline
        Transformer-4       & 3.29M & \textcolor{black}{0.099}  & \textcolor{black}{1.48} & \textcolor{black}{2.84G} & \textcolor{black}{2.31} & \textcolor{black}{7.28G} & \textcolor{black}{3.94} & \textcolor{black}{20.91G}  \\
        ExtBiMamba-3        & 2.76M & \textcolor{black}{0.089}  & \textcolor{black}{1.08} & \textcolor{black}{1.82G} & \textcolor{black}{0.745} & \textcolor{black}{3.64G} & \textcolor{black}{0.695} & \textcolor{black}{7.28G}  \\
        ExtBiMamba-4        & 3.64M & \textcolor{black}{0.102}  & \textcolor{black}{1.38} & \textcolor{black}{2.40G} & \textcolor{black}{0.965} & \textcolor{black}{4.80G} & \textcolor{black}{0.914} & \textcolor{black}{9.60G} \\
        \hdashline
        Transformer-6       & 4.86M & \textcolor{black}{0.125} & \textcolor{black}{2.12} & \textcolor{black}{4.22G} & \textcolor{black}{3.32} & \textcolor{black}{10.83G} & \textcolor{black}{5.89} & \textcolor{black}{31.20G} \\
        ExtBiMamba-5        & 4.51M & \textcolor{black}{0.122} & \textcolor{black}{1.71} & \textcolor{black}{2.98G} & \textcolor{black}{1.19} & \textcolor{black}{5.96G} & \textcolor{black}{1.13} & \textcolor{black}{11.92G} \\
        ExtBiMamba-6        & 5.39M & \textcolor{black}{0.141} & \textcolor{black}{2.07} & \textcolor{black}{3.56G} & \textcolor{black}{1.42} & \textcolor{black}{7.12G} & \textcolor{black}{1.35} & \textcolor{black}{14.23G} \\
        \hline
        \multicolumn{7}{l}{\textcolor{black}{\textbf{Mamba vs. Conformer}}} \\
        \hline
        Conformer-4       & 6.22M & \textcolor{black}{0.160} & \textcolor{black}{2.27} & \textcolor{black}{4.67G} & \textcolor{black}{2.91}  & \textcolor{black}{10.93G}  & \textcolor{black}{4.50} & \textcolor{black}{28.21G} \\
        ExtBiMamba-6      & 5.39M & \textcolor{black}{0.141} & \textcolor{black}{2.07} & \textcolor{black}{3.56G} & \textcolor{black}{1.42}  & \textcolor{black}{7.12G} & \textcolor{black}{1.35} & \textcolor{black}{14.23G}  \\
        ExtBiMamba-7      & 6.26M & \textcolor{black}{0.156} & \textcolor{black}{2.37} & \textcolor{black}{4.14G} & \textcolor{black}{1.64}  & \textcolor{black}{8.28G} & \textcolor{black}{1.57} & \textcolor{black}{16.55G}  \\
        \hdashline
        Conformer-6       & 9.26M & \textcolor{black}{0.215} & \textcolor{black}{3.27} & \textcolor{black}{6.96G} & \textcolor{black}{4.28}  & \textcolor{black}{16.31G} & \textcolor{black}{6.74} & \textcolor{black}{42.16G} \\
        ExtBiMamba-10     & 8.89M & \textcolor{black}{0.206} & \textcolor{black}{3.32} & \textcolor{black}{5.88G} & \textcolor{black}{2.31}  & \textcolor{black}{11.75G}  & \textcolor{black}{2.22} & \textcolor{black}{23.50G}  \\
        \toprule[1.0pt]
    \end{tabular}
    }
    \vspace{-1.0em}
\end{table}

Table~\ref{tab:complexity} presents the comparison results of 
ExtBiMamba with non-causal Transformer and Conformer in training time, RTF, and MACs. We employ different input speech lengths (i.e., 10s, 20s, and 40s) for a comprehensive evaluation. We can find that ExtBiMamba consistently demonstrates lower RTF and MACs Transformer at similar model sizes, and its superiority further expands with longer inference lengths. For 10s and 40s inference lengths, ExtBiMamba-4 (3.64M) achieves RTF values of $0.93\times$ and $0.23\times$, and MACs of $0.85\times$ and $0.46\times$, respectively, versus Transformer-4 (3.29M). Compared to Conformer, a trend similar to the results of the comparison with Transformer is observed. ExtBiMamba has a marginally higher RTF for 10s inference length but significantly lower RTF and MACs for 20s and 40s lengths.

\begin{table}[!t]
    \centering
    % \footnotesize
    % \scriptsize
    \small
    \def\arraystretch{1.3}
    \setlength{\tabcolsep}{1pt}
    \setlength{\abovetopsep}{0pt}
    \setlength\belowbottomsep{0pt} 
    \setlength\aboverulesep{0pt} 
    \setlength\belowrulesep{0pt}
    % \caption{\textcolor{black}{Evaluation results for replacing the MHSA with Mamba/BiMamba layer in Transformer network, across causal and noncausal configurations.}}
    \caption{\textcolor{black}{Evaluation results for replacing the MHSA module with Mamba/BiMamba layer in Transformer and Conformer networks, across causal and non-causal configurations.}}
    \label{tab:transmamba}
    \resizebox{\columnwidth}{!}{
    \begin{tabular}{@{}lcccccccc@{}}
        \toprule[1.0pt]
        \multirow{2}{*}{\textbf{Model}} & \multirow{2}{*}{\textbf{\#Para.}} & \multirow{2}{*}{\textbf{Cau.}} & \multicolumn{6}{c}{\textbf{Metrics}}\\
        \cline{4-9}
        & & & \textcolor{black}{N-PESQ$\uparrow$} & \textcolor{black}{W-PESQ$\uparrow$} & \textcolor{black}{ESTOI$\uparrow$} & \textcolor{black}{CSIG$\uparrow$} & \textcolor{black}{CBAK$\uparrow$} & \textcolor{black}{COVL$\uparrow$}\\
        \midrule
        \multicolumn{9}{l}{\textcolor{black}{\small \textbf{Mamba vs. MHSA in Transformer}}} \\
        \midrule
        Noisy  & -- & -- & \textcolor{black}{1.88} & \textcolor{black}{1.24} & \textcolor{black}{56.12} & \textcolor{black}{2.26} & \textcolor{black}{1.80} & \textcolor{black}{1.67} \\
        Transformer-4 &  3.29M  & \ding{52}   & \textcolor{black}{2.56} & \textcolor{black}{1.84} & \textcolor{black}{70.32} & \textcolor{black}{3.17} & \textcolor{black}{2.39} & \textcolor{black}{2.47} \\
        
        \, \textcolor{black}{MHSA$\rightarrow$Mamba} &  3.99M  & \ding{52}   & \textcolor{black}{\bf 2.65} & \textcolor{black}{\bf 1.93} & \textcolor{black}{\bf 72.28} & \textcolor{black}{\bf 3.24} & \textcolor{black}{\bf 2.45} & \textcolor{black}{\bf 2.55} \\
        
        \hdashline
        Transformer-4 & 3.29M  & \ding{55} & \textcolor{black}{2.74} & \textcolor{black}{2.01} & \textcolor{black}{73.44} & \textcolor{black}{3.31} & \textcolor{black}{2.50} & \textcolor{black}{2.63}  \\
        
        \, \textcolor{black}{MHSA$\rightarrow$InnBiMamba}  &  4.17M  & \ding{55} & \textcolor{black}{2.86} & \textcolor{black}{2.15} & \textcolor{black}{76.24} & \textcolor{black}{3.44} & \textcolor{black}{2.61} & \textcolor{black}{2.78}  \\
        
        \, \textcolor{black}{MHSA$\rightarrow$ExtBiMamba}  &  5.74M  & \ding{55} & \textcolor{black}{\bf 2.88} & \textcolor{black}{\bf 2.18} & \textcolor{black}{\bf 76.75} & \textcolor{black}{\bf 3.54} & \textcolor{black}{\bf 2.63} & \textcolor{black}{\bf 2.84}  \\
        
        \hline
        Transformer-6 &  4.86M  & \ding{52}  & \textcolor{black}{2.60} & \textcolor{black}{1.87} & \textcolor{black}{71.37} & \textcolor{black}{3.20} & \textcolor{black}{2.41} & \textcolor{black}{2.50} \\
        
        \, \textcolor{black}{MHSA$\rightarrow$Mamba} &  5.92M  & \ding{52}  & \textcolor{black}{\bf 2.68} & \textcolor{black}{\bf 1.95} & \textcolor{black}{\bf 73.04} & \textcolor{black}{\bf 3.31} & \textcolor{black}{\bf 2.48} & \textcolor{black}{\bf 2.60} \\
        
        \hdashline
        Transformer-6 & 4.86M  & \ding{55} & \textcolor{black}{2.78} & \textcolor{black}{2.05} & \textcolor{black}{74.56} & \textcolor{black}{3.38} & \textcolor{black}{2.52} & \textcolor{black}{2.69} \\
        
        \, \textcolor{black}{MHSA$\rightarrow$InnBiMamba}  &  6.19M  & \ding{55} & \textcolor{black}{2.89} & \textcolor{black}{2.17} & \textcolor{black}{77.22} & \textcolor{black}{3.49} & \textcolor{black}{2.61} & \textcolor{black}{2.80}  \\
        
        \, \textcolor{black}{MHSA$\rightarrow$ExtBiMamba}  &  8.54M  & \ding{55} & \textbf{2.91} & \textbf{2.21} & \textbf{77.60} & \textbf{3.60} & \textbf{2.65} & \textbf{2.88}  \\

        \toprule
         \multicolumn{9}{l}{\textcolor{black}{\small \textbf{Mamba vs. MHSA in Conformer}}} \\
        \toprule
        Conformer-4 &  6.22M  &  \ding{52}   & \textcolor{black}{2.67} & \textcolor{black}{1.94} & \textcolor{black}{72.78} & \textcolor{black}{\bf 3.29} & \textcolor{black}{2.45} & \textcolor{black}{2.58}  \\
        \, \textcolor{black}{MHSA$\rightarrow$Mamba} &  6.92M & \ding{52}   & \textcolor{black}{\bf 2.69} & \textcolor{black}{\bf 1.95} & \textcolor{black}{\bf 73.27} & \textcolor{black}{\bf 3.29} & \textcolor{black}{\bf 2.48} & \textcolor{black}{\bf 2.59}  \\
        \hdashline
        Conformer-4 & 6.22M  & \ding{55} & \textcolor{black}{2.88} & \textcolor{black}{2.17} & \textcolor{black}{76.68} & \textcolor{black}{3.51} & \textcolor{black}{2.61} & \textcolor{black}{2.82}  \\
        \, \textcolor{black}{MHSA$\rightarrow$InnBiMamba}  &  7.10M  & \ding{55} & 2.89 & 2.17 & 77.40 & 3.51 & 2.62 & 2.81  \\
        \, \textcolor{black}{MHSA$\rightarrow$ExtBiMamba}  &  8.67M  & \ding{55} & \bf 2.92 & \bf 2.20 & \bf 77.74 & \bf 3.57 & \bf 2.64 & \bf 2.88  \\
        \hline
        % \hline
        Conformer-6 &  9.26M  &  \ding{52}   & \textcolor{black}{2.68} & \textcolor{black}{1.94} & \textcolor{black}{73.41} & \textcolor{black}{3.30} & \textcolor{black}{2.46} & \textcolor{black}{2.59}  \\
        \, \textcolor{black}{MHSA $\rightarrow$Mamba} & 10.32M  &  \ding{52}   & \textcolor{black}{\bf 2.71} & \textcolor{black}{\bf 1.95} & \textcolor{black}{\bf 73.69} & \textcolor{black}{\bf 3.34} & \textcolor{black}{\bf 2.49} & \textcolor{black}{\bf 2.62}  \\
        \hdashline
        Conformer-6 & 9.26M  & \ding{55} & \textcolor{black}{2.91} & \textcolor{black}{2.20} & \textcolor{black}{77.56} & \textcolor{black}{3.54} & \textcolor{black}{2.62} & \textcolor{black}{2.84}  \\
        \, \textcolor{black}{MHSA$\rightarrow$InnBiMamba} & 10.59M  & \ding{55} & 2.92 & 2.19 & 77.85 & 3.52 & 2.64 & 2.83 \\
        \, \textcolor{black}{MHSA$\rightarrow$ExtBiMamba} & 12.94M  & \ding{55} & \bf 2.93 & \bf 2.23 & \bf 78.21 & \bf 3.60 & \bf 2.65 & \bf 2.89 \\
        \toprule[1.0pt]
    \end{tabular}
    }
    \vspace{-0.7em}
\end{table}

\textbf{Mamba vs. MHSA}. We also explore replacing the MHSA with the Mamba layer in Transformer and Conformer. Tables~\ref{tab:transmamba} presents the evaluation results of replacing the MHSA module in Transformer and Conformer with the Mamba (or BiMamba) layer. It can be observed that Transformers substantially benefit from using the Mamba and BiMamba layers. TransMamba-4 and TransExtBiMamba-6 improve over causal Transformer-4 and non-causal Transformer-6 by 0.09 and 0.13 in N-PESQ, 0.09 and 0.16 in W-PESQ, 1.96\% and 4.04\% in ESTOI, 0.07 and 0.22 in CSIG, 0.06 and 0.13 in CBAK, and 0.08 and 0.19 in COVL, respectively. In addition, TransMamba-4 (3.99M) and TransInnBiMamba-4 (4.17M) outperform causal Transformer-6 (4.86M) and non-causal Transformer (4.86M), respectively, which further confirms the superiority of Mamba layer over MHSA. Among InnBiMamba and ExtBiMamba, we observe that TransExtBiMamba performs slightly better than TransInnBiMamba. With fewer parameters, TransExtBiMamba-4 (5.74M) achieves slightly higher scores in W-PESQ, CSIG, CBAK, and COVL, but slightly lower scores in N-PESQ and ESTOI than TransInnBiMamba-6 (6.19M).

\begin{table}[!t]
    \centering
    \scriptsize
    \def\arraystretch{1.3}
    \setlength{\tabcolsep}{1pt}
    \setlength{\abovetopsep}{0pt}
    \setlength\belowbottomsep{0pt} 
    \setlength\aboverulesep{0pt} 
    \setlength\belowrulesep{0pt}
    % \caption{\textcolor{black}{Training speed and inference speed of Transformer and TransBiMamba, and Conformer and ConBiMamba.}}
    \caption{\textcolor{black}{Training time, RTF ($\times 10^{-4}$), and MACs of non-causal Transformer and TransBiMamba, and non-causal Conformer and ConBiMamba, across different input speech lengths (10s, 20s, and 40s).}}
    \label{tab:conmambatime}
    \resizebox{\columnwidth}{!}{
    % \begin{tabular}{@{}lccc||ccc@{}}
    %     \toprule[1.0pt]
    %        & Transformer-6  & \textcolor{black}{TransInnBiMamba-6} & \textcolor{black}{TransExtBiMamba-6} & Conformer-6 & \textcolor{black}{ConInnBiMamba-6} & \textcolor{black}{ConExtBiMamba-6} \\
    %       \hline
    %       sec/step & 0.125 & 0.155 & 0.161 & 0.215 & 0.206 & 0.214 \\
    %       RTF & $2.12\times 10^{-4}$ & $1.81\times 10^{-4}$ & $1.95\times 10^{-4}$ &  
    %       $3.27\times 10^{-4}$  & $2.85\times 10^{-4}$ & $2.99\times 10^{-4}$  \\
    %       \textcolor{black}{MACs} & \textcolor{black}{4.22G}  & \textcolor{black}{4.06G}  & \textcolor{black}{5.53G}  & \textcolor{black}{6.96G}  & \textcolor{black}{6.80G}  & \textcolor{black}{8.26G}  \\
    %     \toprule[1.0pt]
    % \end{tabular}

    \begin{tabular}{@{}lccccccccc@{}}
    \toprule[1.0pt]
        \multirow{2}{*}{\textbf{Model}} & \multirow{2}{*}{\textbf{\#Para.}} & \multirow{2}{*}{\textcolor{black}{sec/step$\downarrow$}} & \multicolumn{2}{c}{\textcolor{black}{10s}} & \multicolumn{2}{c}{\textcolor{black}{20s}} & \multicolumn{2}{c}{\textcolor{black}{40s}} \\
        & & & \textcolor{black}{RTF$\downarrow$} & \textcolor{black}{MACs$\downarrow$} & \textcolor{black}{RTF$\downarrow$} & \textcolor{black}{MACs$\downarrow$} & \textcolor{black}{RTF$\downarrow$} & \textcolor{black}{MACs$\downarrow$} \\  
        \hline
        Transformer-6           & 4.86M & \textcolor{black}{0.125} & \textcolor{black}{2.12} & \textcolor{black}{4.22G} & \textcolor{black}{3.32} & \textcolor{black}{10.83G} & \textcolor{black}{5.89} & \textcolor{black}{31.20G} \\
        
        TransInnBiMamba-6        & 6.19M & \textcolor{black}{0.155} & \textcolor{black}{1.81} & \textcolor{black}{4.06G} & \textcolor{black}{1.63} & \textcolor{black}{8.11G} & \textcolor{black}{1.55} & \textcolor{black}{16.22G} \\
        
        TransExtBiMamba-6        & 8.54M & \textcolor{black}{0.161} & \textcolor{black}{1.95} & \textcolor{black}{5.53G} & \textcolor{black}{1.76} & \textcolor{black}{11.05G} & \textcolor{black}{1.69} & \textcolor{black}{22.10G} \\
        \hline
        \hline
        Conformer-6       & 9.26M & \textcolor{black}{0.215} & \textcolor{black}{3.27} & \textcolor{black}{6.96G} & \textcolor{black}{4.28}  & \textcolor{black}{16.31G} & \textcolor{black}{6.74} & \textcolor{black}{42.16G} \\
        
        ConInnBiMamba-6  & 10.59M & \textcolor{black}{0.206} & \textcolor{black}{2.85} & \textcolor{black}{6.80G} & \textcolor{black}{2.56}  & \textcolor{black}{11.75G}  & \textcolor{black}{2.45} & \textcolor{black}{23.50G}  \\
        
        ConExtBiMamba-6   & 12.94M & \textcolor{black}{0.214} & \textcolor{black}{2.99} & \textcolor{black}{8.26G} & \textcolor{black}{2.72}  & \textcolor{black}{16.53G}  & \textcolor{black}{2.60} & \textcolor{black}{33.06G}  \\
    \toprule[1.0pt]
    \end{tabular}}
    \vspace{-1.5em}
\end{table}

% TransInnBiMamba-6           1.8127/1.6293/1.5539
% TransExtBiMamba-6           1.8879/1.7584/1.6906

% ConInnBiMamba-6             2.80/2.5606/2.4505
% ConExtBiMamba-6             2.91/2.7169/2.5987

% ConExtBiMamba-6           8.26G/16.53G/33.06G
% ConInnBiMamba-6           6.795G/13.591G/27.182G
% TransExtBiMamba-6         5.526G/11.051G/22.102G
% TransInnBiMamba-6         4.056G/8.112G/16.223G
% InnBiMamba-9              3.091G/6.182G/12.365G
% InnBiMamba-13             4.428G/8.857G/17.713G

% Transformer-4             2.842G/7.275G/20.908G
% Transformer-6             4.222G/10.829G/31.197G
% Conformer-4               4.669G/10.927G/28.214G
% Conformer-6               6.962G/16.307G/42.156G
% BiMamba-3                 1.820G/3.641G/7.282G
% BiMamba-4                 2.400G/4.799G/9.599G
% BiMamba-5                 2.979G/5.958G/11.916G
% BiMamba-6                 3.558G/7.116G/14.233G
% BiMamba-7                 4.137G/8.275G/16.550G
% BiMamba-8                 4.717G/9.433G/18.867G
% BiMamba-10                5.875G/11.750G/23.501G
% ConExtBiMamba-6           8.26G/16.53G/33.06G
% ConInnBiMamba-6           6.795G/13.591G/27.182G
% TransExtBiMamba-6         5.526G/11.051G/22.102G
% TransInnBiMamba-6         4.056G/8.112G/16.223G
% InnBiMamba-9              3.091G/6.182G/12.365G
% InnBiMamba-13             4.428G/8.857G/17.713G

% Transformer-4             2.842G/7.275G/20.908G
% Transformer-6             4.222G/10.829G/31.197G
% Conformer-4               4.669G/10.927G/28.214G
% Conformer-6               6.962G/16.307G/42.156G

As shown in Table~\ref{tab:transmamba}, we observe that the Conformer architecture can benefit from the use of Mamba and ExtBiMamba. The ConExtBiMamba-4 and ConExtBiMamba-6 outperform non-causal Conformer-4 and Conformer-6 by 0.04 and 0.02 in N-PESQ, 0.03 and 0.03 in W-PESQ, 1.06\% and 0.65\% in ESTOI, 0.06 and 0.06 in CSIG, 0.03 and 0.03 in CBAK, and 0.06 and 0.05 in COVL, respectively. In addition, ConExtBiMamba-4 (8.67M) also exhibits a slightly better performance than non-causal Conformer-6 (9.26M). Among InnBiMamba and ExtBiMamba, similarly, ConExtBiMamba performs better than ConInnBiMamba. Table~\ref{tab:conmambatime} compares non-causal Transformer and TransBiMamba, as well as non-causal Conformer and ConBiMamba, in terms of training time, RTF, and MACs, across inference lengths of 10s, 20s, and 40s. We can observe that the replacement of MHSA with both InnBiMamba and ExtBiMamba leads to lower RTF, particularly for longer input speech lengths.
\begin{table}[!t]
% \color{blue}
\centering
    \scriptsize
    \def\arraystretch{1.05}
     \setlength{\abovetopsep}{0pt}
    \setlength\belowbottomsep{0pt} 
    \setlength\aboverulesep{0pt} 
    \setlength\belowrulesep{0pt}
    \setlength{\tabcolsep}{0.6pt}
\caption{\textcolor{black}{The comparison results of the models on VoiceBank+DEMAND benchmark dataset. $\dagger$ denotes the results reproduced using the source code provided by the authors.}}
% \vspace*{0.1in}
% \scalebox{1.18}
\resizebox{\columnwidth}{!}{%
{\color{black}\begin{tabular}{llccccc }
% \hline
\toprule
\textbf{Method}  & \textbf{\#Para.} & W-PESQ$\uparrow$  & STOI$\uparrow$ & CSIG$\uparrow$  & CBAK$\uparrow$  & COVL$\uparrow$  \\
\hline 
Noisy & - & 1.97 & 0.92 & 3.35 & 2.44 & 2.63 \\
% \hdashline
\hline 
% \hline
SEGAN~\cite{SEGAN} & 43.2M       
& 2.16 & 0.93 & 3.48 & 2.94 & 2.80  \\
DSEGAN~\cite{DSEGAN} & --       
& 2.35 & 0.93 & 3.55 & 3.10 & 2.93  \\
WaveCRN~\cite{wavecrn} & 4.66M   
& 2.64 & -- & 3.94 & 3.37 & 3.29  \\
MHSA+SPK~\cite{koizumi2020speech} & -- 
& 2.99 & -- & 4.15 & 3.42 & 3.57 \\
HiFi-GAN~\cite{hifigan} & -- 
& 2.94 & -- & 4.07 & 3.07 & 3.49 \\
% \textcolor{black}{Wave-U-Net \cite{}} & 10.2M       
% & 2.40 & -- & 3.52 & 3.24 & 2.96  \\
MetricGAN~\cite{metricgan} & 1.89M   
& 2.86 & -- & 3.99 & 3.18 & 3.42  \\
PHASEN~\cite{phasen} & 8.41M   
& 2.99 & -- & 4.21 & 3.55 & 3.62  \\
TFT-Net\cite{tftnet} & --   
& 2.75 & -- & 3.93 & 3.44 & 3.34  \\
DCCRN~\cite{dccrn} & 3.7M   
& 2.68 & 0.94 & 3.88 & 3.18 & 3.27  \\
DCCRN+~\cite{dccrn+}& 3.3M   
& 2.84 & -- & -- & -- & --  \\
S-DCCRN~\cite{sdccrn} & 2.34M   
& 2.84 & 0.94 & 4.03 & 2.97 & 3.43  \\
SADNUNet~\cite{sadnunet} & 2.63M 
& 2.82 & 0.95 & 4.18 & 3.47 & 3.51 \\
SEAMNET~\cite{seamnet} & 5.1M   
& -- & -- & 3.87 & 3.16 & 3.23  \\
SA-TCN~\cite{sa-tcn} & 3.76M   
& 2.99 & 0.94 & 4.25 & 3.45 & 3.62  \\
DeepMMSE~\cite{deepmmse} & 1.98M   
& 2.95 & 0.94 & 4.28 & 3.46 & 3.64  \\
DCTCN~\cite{dctcn} & 9.7M 
& 2.83 & -- & 3.91 & 3.37 & 3.37 \\
CleanUNet~\cite{cleanunet} & 46.07M 
& 2.91 & \textbf{0.96} & 4.34 & 3.42 & 3.65 \\
SE-Conformer~\cite{conformer-se} & --   
& 3.13 & 0.95 & 4.45 & 3.55 & 3.82  \\
MetricGAN+~\cite{metricgan+} & --   
& 3.15 & -- & 4.14 & 3.16 & 3.64  \\
T-GSA~\cite{tgsa} & --   
& 3.06 & -- & 4.18 & 3.59 & 3.62  \\
DEMCUS~\cite{demcus} & 58M  
& 3.07 & 0.95 & 4.31 & 3.40 & 3.63  \\
SGMSE+~\cite{segmse} & --   
& 2.96 & -- & -- & -- & --  \\
StoRM~\cite{storm} & 27.8M   
& 2.93 & -- & -- & -- & --  \\
SGMSE+M~\cite{storm} & --   
& 2.96 & -- & -- & -- & --  \\
ResTCN+TFA-Xi~\cite{tfaj} & 1.98M   
& 3.02 & 0.94 & 4.32 & 3.52 & 3.68  \\
CMGAN~\cite{cmgan} & 1.83M & 3.41 & 0.96 & 4.63 & 3.94 & 4.12 \\
MP-SENet-Conformer$^\dagger$~\cite{mpsenet} & 2.05M   
& 3.46 & \bf 0.96 & 4.66 & 3.92 & 4.15   \\
MP-SENet-SA+BiGRU$^\dagger$~\cite{lu2023explicit} & 2.26M & \bf 3.53& \bf \bf 0.96& \bf 4.72& \bf 3.95& \bf 4.25  \\
MP-SENet-ExtBiMamba & 2.19M & 3.48 &  \bf 0.96 &  4.69 & 3.94 &  4.17   \\
SEMamba (-CL)$^\dagger$~\cite{semamba} & 2.33M& 3.51& \bf 0.96& 4.70& 3.93& 4.19\\
% \hline
% cmgan
\toprule
\end{tabular}
}}
\label{demand}
\vspace{-1.5em}
\end{table}

\textbf{VoiceBank+DEMAND Benchmark.} In Table~\ref{demand}, we evaluate the use of ExtBiMamba on the VoiceBank+DEMAND dataset~\cite{demand}, which is a commonly used benchmark for speech enhancement. Results regarding \cite{semamba, mpsenet, lu2023explicit} are obtained using the source code provided by the authors, with their respective default parameter configurations. To ensure a fair comparison, we follow the model structures described in the referenced works and reproduce their results. We denote the MP-SENet models proposed in \cite{mpsenet} and \cite{lu2023explicit} as MP-SENet-Conformer and MP-SENet-SA+BiGRU, respectively, reflecting their architectural differences. For SEMamba~\cite{semamba}, we exclude the use of the consistency loss (CL) and perceptual contrast stretching (PCS), as these components are not utilized in the other baseline methods and may introduce an unfair advantage. In addition, MP-SENet-Conformer involves relatively complex model structure designs, such as the two-stage Conformer backbone network, magnitude mask and phase decoders, PSEQ-based metric generative adversarial training, and joint optimization of multiple loss functions. Hence, the independent ExtBiMamba model is not directly comparable to MP-SENet-Conformer. We thus substitute the Conformer module within MP-SENet-Conformer with our proposed ExtBiMamba, resulting in MP-SENet-ExtBiMamba. Although MP-SENet-ExtBiMamba does not achieve the highest performance among all approaches, the experimental results clearly demonstrate that replacing the Conformer with ExtBiMamba yields consistent performance improvements.

\begin{table*}[t] 
\centering
\scriptsize
\def\arraystretch{1.25}
% 增大列间距
\setlength{\tabcolsep}{4pt}
\setlength{\abovetopsep}{0pt}
\setlength{\belowbottomsep}{0pt} 
\setlength{\aboverulesep}{0pt} 
\setlength{\belowrulesep}{0pt}
\caption{ASR results for the different datasets without language model in WER/MER (\%). ``ESPnet'' indicates the best result reported by ESPnet or related works, where the SEAME and ASRU numbers are taken from~\cite{liu23_icassp,liu2024enhancing, liu2024aligning}.}
\resizebox{0.95\textwidth}{!}{%
\begin{tabular}{@{}l@{\hspace{4pt}}*{12}{c}@{}}
\toprule[1.0pt]
\multirow{2}{*}{\textbf{Method}} 
  & \multicolumn{3}{c}{LibriSpeech-100} 
  & \multicolumn{3}{c}{LibriSpeech-960} 
  & \multicolumn{3}{c}{SEAME} 
  & \multicolumn{3}{c}{ASRU} \\
\cmidrule(lr){2-4} \cmidrule(lr){5-7} \cmidrule(lr){8-10} \cmidrule(lr){11-13}
 & \textcolor{black}{\#Para.} & dev\textdownarrow & test\textdownarrow
  & \textcolor{black}{\#Para.} & dev\textdownarrow & test\textdownarrow 
  & \textcolor{black}{\#Para.} & man\textdownarrow & sge\textdownarrow 
  & \textcolor{black}{\#Para.} & dev\textdownarrow & test\textdownarrow \\
\midrule
\textit{ESPnet} 
  &    &  &  &    &  &  &    &  &  &   &  &  \\[0.5mm]
\quad Conformer~\cite{gulati20_interspeech} 
  & \textcolor{black}{34.23M}   & 6.3 & 6.5 
  & \textcolor{black}{116.15M}   & 2.1 & 2.4 
  & \textcolor{black}{--}   & 16.6 & 23.3 
  & \textcolor{black}{--}   & --  & 12.2 \\[0.5mm]
\quad Branchformer~\cite{peng2022branchformer} 
  & \textcolor{black}{38.96M}   & 6.1 & 6.3 
  & \textcolor{black}{126.24M}   & 2.2 & 2.4 
  & \textcolor{black}{--}   & --  & -- 
  & \textcolor{black}{--}   & --  & -- \\
\hline\hline
\multicolumn{13}{l}{\textit{Reproduced Results}} \\
\quad Branchformer~\cite{peng2022branchformer} 
  & \textcolor{black}{38.96M}   & 6.3 & 6.4 
  & \textcolor{black}{126.25M}   & 2.2 & 2.4 
  & \textcolor{black}{38.96M}   & \textbf{16.3} & \textbf{23.2} 
  & \textcolor{black}{38.96M}   & 12.5 & 11.8 \\
\hdashline
\quad Mamba 
  & \textcolor{black}{20.41M}   & 40.8 & 40.0 
  & \textcolor{black}{93.02M}   & 21.8 & 22.3 
  & \textcolor{black}{20.41M}   & 44.5 & 55.3 
  & \textcolor{black}{20.41M}   & 38.0 & 36.3 \\
\quad InnBiMamba 
  & \textcolor{black}{20.94M}   & 39.6 & 38.2 
  & \textcolor{black}{93.52M}   & 21.8 & 22.5 
  & \textcolor{black}{20.94M}   & 44.3 & 55.4 
  & \textcolor{black}{20.94M}   & 38.4 & 37.9 \\
\quad ExtBiMamba 
  & \textcolor{black}{25.79M}   & 38.5 & 37.7 
  & \textcolor{black}{98.20M}   & 21.6 & 22.1 
  & \textcolor{black}{25.79M}   & 44.4 & 55.2 
  & \textcolor{black}{25.79M}   & 38.2 & 36.8 \\
\hdashline
\quad Transformer~\cite{vaswani2017attention} 
  & \textcolor{black}{29.38M} & 8.0 & 8.4 
  & \textcolor{black}{99.36M} & 2.8 & 3.2 
  & \textcolor{black}{29.86M} & 17.7 & 24.5 
  & \textcolor{black}{30.86M} & 13.7 & 13.1 \\
\quad \textcolor{black} {\,\, MHSA$\rightarrow$Mamba }
  & \textcolor{black}{32.53M}   & 10.9 & 11.2 
  & \textcolor{black}{102.51M}   & 3.2 & 3.5 
  & \textcolor{black}{34.06M}   & 20.7 & 29.5 
  & \textcolor{black}{34.01M}   & 24.2 & 23.1 \\
\quad \textcolor{black} {\,\, MHSA$\rightarrow$InnBiMamba }
  & \textcolor{black}{33.34M} & 8.8 & 9.4 
  & \textcolor{black}{103.35M} & 2.5 & 3.0 
  & \textcolor{black}{35.14M} & 18.4 & 26.0 
  & \textcolor{black}{34.82M} & 20.2 & 19.5 \\
\quad \textcolor{black} {\,\, MHSA$\rightarrow$ExtBiMamba }
  & \textcolor{black}{40.42M} & 8.4 & 8.7 
  & \textcolor{black}{110.04M} & 2.5 & 2.8 
  & \textcolor{black}{44.56M} & 17.2 & 24.3 
  & \textcolor{black}{41.89M} & 18.7 & 18.0 \\
\hdashline
\quad Conformer~\cite{gulati20_interspeech} 
  & \textcolor{black}{34.23M} & 6.3 & 6.5 
  & \textcolor{black}{116.15M} & 2.3 & 2.6 
  & \textcolor{black}{47.27M} & 16.9 & 23.6 
  & \textcolor{black}{48.27M} & 12.8 & 12.2 \\
\quad \textcolor{black} {\,\, MHSA$\rightarrow$Mamba }
  & \textcolor{black}{36.33M}   & 6.6 & 6.9 
  & \textcolor{black}{119.30M}   & 2.6 & 2.9 
  & \textcolor{black}{49.37M}   & 17.7 & 24.9 
  & \textcolor{black}{50.37M}   & 13.5 & 12.9 \\
  \quad \textcolor{black} {\,\, MHSA$\rightarrow$InnBiMamba }
  & \textcolor{black}{36.89M} & 6.0 & 6.4 
  & \textcolor{black}{120.11M} & 2.1 & 2.3 
  & \textcolor{black}{49.91M} & 17.1 & 23.8 
  & \textcolor{black}{50.91M} & 12.7 & 12.2 \\
\quad \textcolor{black} {\,\, MHSA$\rightarrow$ExtBiMamba }
  & \textcolor{black}{41.59M} & \textbf{5.9} & \textbf{6.0} 
  & \textcolor{black}{123.51M} & \textbf{2.0} & \textbf{2.3} 
  & \textcolor{black}{54.62M} & 16.6 & 23.4 
  & \textcolor{black}{55.62M} & \textbf{12.3} & \textbf{11.5} \\
\toprule[1.0pt]
\end{tabular}%
}
\label{ASR Main Result}
\end{table*}

\begin{table}[t]
\centering
% \scriptsize
\small
\def\arraystretch{1.25}
\setlength{\tabcolsep}{2pt} % 减少列间距
\setlength{\abovetopsep}{0pt}
\setlength\belowbottomsep{0pt} 
\setlength\aboverulesep{0pt} 
\setlength\belowrulesep{0pt}
\caption{WER results on LibriSpeech (without external language model) across various parameter counts and frameworks.}
\resizebox{0.95\columnwidth}{!}{ % 缩小表格以适应单列
\begin{tabular}{lccccc}
\toprule[1.25pt]
\textbf{\textcolor{black}{Model}} & \textbf{\#Para.} & \textbf{dev\textdownarrow} & \textbf{dev other\textdownarrow} & \textbf{test\textdownarrow} & \textbf{test other\textdownarrow} \\
\midrule
\multicolumn{4}{l}{\textbf{\textit{LibriSpeech100}}} \\
\textcolor{black}{Mamba} & \textcolor{black}{20.41M} & \textcolor{black}{40.8} & - & \textcolor{black}{40.0} & - \\
\textcolor{black}{Mamba Large} & \textcolor{black}{32.52M} & \textcolor{black}{38.8} & - & \textcolor{black}{39.2} & - \\
\textcolor{black}{ExtBiMamba} & \textcolor{black}{25.79M} & \textcolor{black}{38.5} & - & \textcolor{black}{37.7} & - \\
\textcolor{black}{ExtBiMamba Large} & \textcolor{black}{33.52M} & \textcolor{black}{37.8} & - & \textcolor{black}{37.2} & - \\
\hdashline
Transformer~\cite{vaswani2017attention} & 29.38M & 8.0 & 20.0 & 8.4 & 20.2 \\
\, \textcolor{black}{\, MHSA$\rightarrow$InnBiMamba} & 33.34M & 8.8 & 23.3 & 9.4 & 23.6 \\
\, \textcolor{black}{ \, MHSA$\rightarrow$ExtBiMamba }& 40.42M & 8.4 & 22.4 & 8.7 & 23.1 \\
\hdashline
Conformer Large & 42.17M & 6.2 & 17.4 & 6.4 & 17.3 \\
Conformer~\cite{gulati20_interspeech} & 34.23M & 6.3 & 17.4 & 6.5 & 17.3 \\
\, \textcolor{black}{\, MHSA$\rightarrow$InnBiMamba} & 36.89M & 6.0 & 17.4 & 6.4 & 17.6 \\
\, \textcolor{black}{\, MHSA$\rightarrow$InnBiMamba Large} & 49.90M & 6.1 & 17.3 & 6.3 & 17.3 \\
\, \textcolor{black}{\, MHSA$\rightarrow$ExtBiMamba} & 41.59M & \textbf{5.9} & \textbf{17.1} & \textbf{6.0} & \textbf{17.2} \\
\midrule
\multicolumn{4}{l}{\textbf{\textit{LibriSpeech960}}} \\
Transformer~\cite{vaswani2017attention} & 99.36M & 2.8 & 7.6 & 3.2 & 7.5 \\
\, \textcolor{black}{\, MHSA$\rightarrow$InnBiMamba}& 103.35M & 2.5 & 7.4 & 3.0 & 7.3 \\
\, \textcolor{black}{\, MHSA$\rightarrow$ExtBiMamba}& 110.04M & 2.5 & 6.9 & 2.8 & 6.9 \\
\hdashline
Conformer~\cite{gulati20_interspeech} & 116.15M & 2.3 & 5.5 & 2.6 & 5.6 \\
\quad \textcolor{black}{\, MHSA$\rightarrow$InnBiMamba}& 120.11M & 2.1 & 5.6 & 2.4 & 5.5 \\
\, \textcolor{black}{\, MHSA$\rightarrow$ExtBiMamba}& 123.51M & \textbf{2.0} & \textbf{5.4} & \textbf{2.3} & \textbf{5.4} \\
\toprule[1.0pt]
\end{tabular}
}
\label{Completed Result on liber960}
% \vspace{-1.3em}
\end{table}

\subsection{Speech Recognition}\label{sec:asr_results}
\textbf{Independent vs. Substitute for MHSA.} Table~\ref{ASR Main Result} reports the performance of Mamba when used independently and as a replacement for MHSA across various datasets. We can observe that independent Mamba and BiMamba models exhibit significantly lower performance compared to the Transformer and Conformer models (with the same number of layers). Since the Mamba has fewer parameters when configured with the same number of layers as the Transformer and Conformer, we increased its number of layers as shown in Table~\ref{Completed Result on liber960} to match the number of parameters with the Conformer to further evaluate Mamba for ASR. The performance of Mamba, however, remained undesirable. 

In contrast to the independent Mamba/BiMamba, we observe a significant improvement in the ASR task when Mamba/BiMamba is used as a replacement for MHSA. Specifically, replacing MHSA with ExtBiMamba in the Conformer~(named ConExtBiMamba) exhibits higher performance than the SOTA performance achieved by Conformer and Branchformer across multiple datasets with the same training setups. Additionally, ConExtBiMamba provides faster training and inference speeds compared to the Conformer model as detailed in Table~\ref{tab:training_time_asr}. When we replaced MHSA with ExtBiMamba, the number of parameters exceeded that of the original Conformer. To eliminate the possibility that the performance improvement results from the higher number of parameters, we increase the size of Conformer to a similar number of parameters as ConExtBiMamba, and present the results in Table~\ref{Completed Result on liber960}. We found that although the Conformer's performance improves, it still underperforms ConExtBiMamba and ConInnBiMamba, supporting the effectiveness of our approach.

Our experiments indicate that increasing the number of layers to match parameter sizes does not proportionally improve performance. We can explain it by the rule of diminishing returns~\cite{gupta2018diminishing}. Even when training is stable, a point of diminishing returns~\cite{gupta2018diminishing} is often reached as depth increases. Each additional layer yields progressively smaller improvements once the model’s capacity exceeds the complexity of the task. Essentially, the network hits a complexity ceiling where extra layers learn redundant features or noise. The model's capacity was likely sufficient at moderate depth, so additional layers provided minimal new information. Thus, employing extra layers may even start to hurt if they cause overfitting or optimization issues

Additionally, unlike in the Conformer model, replacing MHSA with Mamba modules in the Transformer does not consistently lead to performance improvements for ASR tasks. Due to the limited inherent nonlinearity of Mamba modules, Mamba-based models require stronger nonlinear capabilities to effectively learn high-level abstractions in speech. The Conformer architecture includes additional convolutional modules with nonlinear activations than the Transformer. These additional components compensate for the weak nonlinearity of the Mamba module, thereby enabling more effective representation learning. This architectural distinction explains why the integration of BiMamba proves more beneficial in the Conformer model than in the Transformer.

Apart from the lower performance than serving as an alternative to MHSA, using Mamba independently poses training stability issues. Due to its weak inherent nonlinearity, independent Mamba may struggle to capture the high-level abstractions required for ASR tasks, consequently hindering stable and effective model training.
\begin{table}[!t]
    \centering
    % \scriptsize % 调整字体大小
    \small
    \def\arraystretch{1.12}
    \setlength{\tabcolsep}{2pt} % 适当减少列间距
    \setlength{\abovetopsep}{0pt}
    \setlength\belowbottomsep{0pt} 
    \setlength\aboverulesep{0pt} 
    \setlength\belowrulesep{0pt}
    \caption{\textcolor{black}{Training time (mins/epoch) and inference speed (RTF) for ASR in LibriSpeech100 on NVIDIA Tesla V100.}}
    \label{tab:training_time_asr}
    % \resizebox{1\columnwidth}{!}{ % 将宽度调整为0.75倍，呈现更小的表格
    \begin{tabular}{@{}lccc@{}}
    \toprule[1.0pt]
          & Conformer & \textcolor{black}{ConInnBiMamba} & \textcolor{black}{ConExtBiMamba} \\
          \hline
          mins/epoch & 21.4 & 18.3 & 19.8 \\
          RTF & 0.179 & 0.174 & 0.177 \\
    \toprule[1.0pt]
    \end{tabular}
    % }
    \vspace{-1.0em}
\end{table}

\textbf{Unidirectional vs. Bidirectional}. Tables~\ref{ASR Main Result}-\ref{Completed Result on liber960} present the evaluation results of unidirectional Mamba and two bidirectional Mamba modules (InnBiMamba and ExtBiMamba) along with their model sizes. Both InnBiMamba and ExtBiMamba show significant improvements over unidirectional Mamba when replacing the MHSA module in Transformer and Conformer frameworks, confirming the effectiveness of bidirectional modeling. Additionally, ExtBiMamba consistently outperforms InnBiMamba across various frameworks and datasets, as demonstrated in Tables~\ref{Completed Result on liber960}. Even when increasing the number of layers in InnBiMamba to match the parameter count of ExtBiMamba for a fair comparison, ExtBiMamba still achieves better performance. Moreover, further comparisons with Conformer-Large and ConInnBiMamba-Large indicate that the performance improvements in ASR are not solely due to an increase in model parameters.

\subsection{Ablation Study on ConExtBiMamba}
We employ the ConExtBiMamba model, which improves the SOTA result achieved by Branchformer in the ASR task, for ablation studies.

\textbf{Hyper-Parameter for BiMamba in ConExtBiMamba}.
The default hyper-parameters for ConExtBiMamba are as follows: the SSM state dimension $N=16$, the state expansion factor $E_{f}=2$, and the local convolution width $d_{\text{conv}}=4$. From Table~\ref{Ablation Study on LibriSpeech100}, we find that increasing the state expansion factor or decreasing the local convolution width has little impact on the performance of ConExtBiMamba. We further investigate the impact of the \(\mathbf{A}\) matrix on performance. The \(\mathbf{A}\) matrix plays a crucial role in both S4 and Mamba. In the original Mamba paper, it is initialized using a real-valued diagonal matrix. However, ensuring randomness is important in deep learning~\cite{koehn2020neural}. To explore this, we examine four types of \(\mathbf{A}\) matrices: a real-valued diagonal matrix, a completely randomized matrix, complex-valued matrix and a real-valued diagonal matrix initialized by multiplying each element with noise generated from a Gaussian distribution. Through our experiments, we discovered that the diagonal matrix with Gaussian noise yielded the best results.
\begin{table}[!t]
\centering
% \scriptsize
\small
\def\arraystretch{1.05}
\setlength{\tabcolsep}{1pt} % 减少列间距
\caption{\textcolor{black}{Ablation Studies for ConExtBiMamba on LibriSpeech100.}}
\resizebox{0.95\columnwidth}{!}{ % 缩小表格以适应单列
\begin{tabular}{lcccc}
\toprule[1.0pt]
\textcolor{black}{\textbf{Model}} & \textbf{dev\textdownarrow} & \textbf{dev other\textdownarrow} & \textbf{test\textdownarrow} & \textbf{test other\textdownarrow} \\
\midrule
\textcolor{black}{ConExtBiMamba w/ Gaussian Noise} & 5.9 & 17.1 & 6.0 & 17.3 \\
% \quad $\rightarrow$ $d_{\text{state}}=32$ & 5.9 & 17.1 & 6.0 & 17.3 \\
\quad $\rightarrow$ \textcolor{black}{$N=32$} & 5.9 & 17.1 & 6.0 & 17.3 \\
\quad $\rightarrow$ $d_{\text{conv}}=2$ & 6.0 & 17.1 & 6.2 & 17.3 \\
% \quad $\rightarrow$ $E=4$ & 5.9 & 17.1 & 6.0 & 17.3 \\
\quad $\rightarrow$ \textcolor{black}{$E_{f}=4$} & 5.9 & 17.1 & 6.0 & 17.3 \\
\quad $\rightarrow$ Random Noise $\mathbf{A}$ matrix & 6.0 & 17.2 & 6.2 & 17.4 \\
\quad $\rightarrow$ default $\mathbf{A}$ matrix & 6.1 & 17.2 & 6.2 & 17.5 \\
\quad \textcolor{black}{$\rightarrow$ complex-valued $\mathbf{A}$ matrix} & \textcolor{black}{6.0} & \textcolor{black}{17.2} & \textcolor{black}{6.2} & \textcolor{black}{17.3} \\
\midrule
\midrule
\textcolor{black}{ConExtBiMamba} & 5.9 & 17.1 & 6.0 & 17.3 \\
\quad $-$ Macaron & 6.1 & 17.7 & 6.3 & 17.9 \\
\quad $-$ Swish & 6.2 & 17.7 & 6.4 & 18.1 \\
\quad $-$ Dropout & 6.0 & 17.1 & 6.0 & 17.4 \\
\quad $+$ Positional Encoding & 6.0 & 17.1 & 6.0 & 17.3 \\
\toprule[1.0pt]
\end{tabular}}
\label{Ablation Study on LibriSpeech100}
\end{table}

\textbf{Hyper-Parameters for ConExtBiMamba}
We conduct ablation experiments on Swish Activation, Macaron-style feed-forward layers, positional encoding, and dropout. From Table~\ref{Ablation Study on LibriSpeech100}, it is apparent that Swish Activation and Macaron-style feed-forward layers enhance the model's performance, while positional encoding and dropout do not have a significant impact. We suggest that the reason positional encoding, a crucial component in Transformer-based models, does not significantly affect the ConExtBiMamba model is twofold. Firstly, Mamba independently models each input, which suggests that the model may better differentiate between different positions. Secondly, because Mamba operates similarly to an RNN, the dependencies between inputs inherently embed some positional information into the model, thus diminishing the impact of positional encoding.
\begin{table}[!htbp]
\centering
% \scriptsize
\small
\def\arraystretch{1.25}
\setlength{\tabcolsep}{2pt} % 减少列间距
    \setlength{\abovetopsep}{0pt}
    \setlength\belowbottomsep{0pt} 
    \setlength\aboverulesep{0pt} 
    \setlength\belowrulesep{0pt}
\caption{Performance on extremely small dataset AN4 by employing WER (\%)}
\resizebox{0.9\columnwidth}{!}{ % 缩小表格以适应单列
\begin{tabular}{lccccccc}
\toprule[1.0pt]
% \textbf{Method/Seed} & \textbf{2048} & \textbf{233} & \textbf{666} & \textbf{1024} & \textbf{3407} & \textbf{Average} & \textbf{Variance}\\
\multirow{2}{*}{\textbf{Model}} & \multicolumn{5}{c}{\textbf{Seed}} & \multirow{2}{*}{\textbf{Average}} & \multirow{2}{*}{\textbf{Variance}} \\
\cline{2-6}
& 2048 & 233 & 666 & 1024 & 3407 &  & \\
\midrule
Conformer & 4.0 & 8.2 & 6.1 & 4.4 & 27.4 & 10.02 & 77.71 \\
\textcolor{black}{ConExtBiMamba} & 3.8 & 3.8 & 4.8 & 3.8 & 6.5 & 4.54 & 1.11 \\
\toprule[1.0pt]
\end{tabular}}
% \vspace{-1.0em}
\label{Performance on Extremely Small Dataset table}
\end{table}

\begin{figure}[!htbp]
    \centering
    \begin{subfigure}[b]{0.48\columnwidth} % 调整子图宽度为单列的一半
        \includegraphics[width=\textwidth]{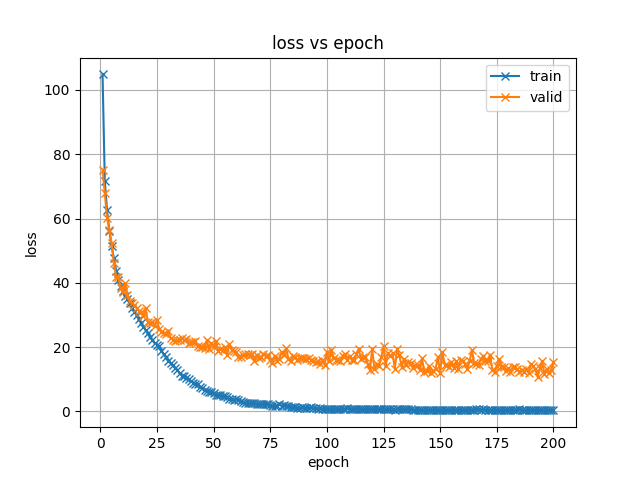}
        \caption{Loss of Conformer}
        \label{conformer loss}
    \end{subfigure}
    \hfill
    \begin{subfigure}[b]{0.48\columnwidth} % 调整子图宽度为单列的一半
        \includegraphics[width=\textwidth]{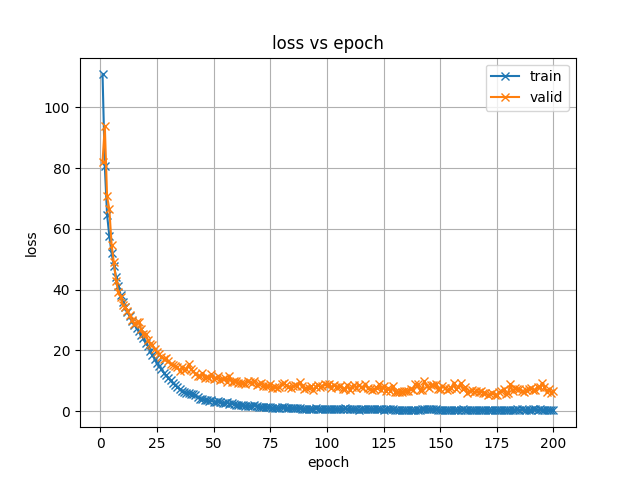}
        \caption{Loss of ConExtBiMamba}
        \label{conbimamba Loss}
    \end{subfigure}
     \caption{\textcolor{black}{Train and Dev Set Loss of Conformer and ConExtBiMamba in AN4 dataset}}
    \label{fig:loss}
\end{figure}

\textbf{Performance on Extremely Small Datasets}
To evaluate model robustness, we conduct tests on the AN4 dataset, a notably small dataset, by averaging the results obtained from five randomly selected seeds for both the Conformer and ConExtBiMamba models under identical hyperparameters. Our results show that ConExtBiMamba consistently outperforms the Conformer across all seeds. Specifically, the mean WER for the Conformer was 10.02, while it was significantly lower for ConExtBiMamba at just 4.54. As shown in Fig.~\ref{conformer loss} and Fig.~\ref{conbimamba Loss}, both models converged properly on the development set, with ConExtBiMamba exhibiting more stable convergence behavior. Remarkably, ConExtBiMamba also demonstrated exceptional robustness, with a WER variance of only 1.11 compared to the Conformer's substantially higher variance of 77.71. The stability in both the loss curves and WER variance suggests that ConExtBiMamba is less prone to overfitting on small datasets compared with the Conformer. This suggests that ConExtBiMamba, like RNNs, benefits from its ability to process information recursively over time, updating only a small set of parameters with each iteration, which helps prevent overfitting in scenarios with limited data. Additionally, this framework also avoids the gradient vanishing and explosion issues typically associated with RNNs. Combining these observations with our main results, we can conclude that ConExtBiMamba effectively merges the strengths of MHSA-based models and RNN models.
\begin{table}[!htbp]
\centering
\def\arraystretch{1.23}
\setlength{\abovetopsep}{0pt}
\setlength\belowbottomsep{0pt} 
\setlength\aboverulesep{0pt} 
\setlength\belowrulesep{0pt}
\caption{Ablation Studies on LibriSpeech100 for Transformer and ExtBiMamba by WER.}
\begin{tabular*}{0.75\columnwidth}{@{\extracolsep{\fill}} lcc}
\toprule[1.0pt]
\textbf{Model}  & \textbf{dev} & \textbf{test} \\
\midrule
Transformer~\cite{vaswani2017attention} & 8.0 & 8.4 \\
\quad $-$ Residual & 45.7 & 54.8 \\
\quad $-$ FFN & 23.2 & 25.4  \\
\hdashline
ExtBiMamba & 38.5 & 37.7 \\
\quad $+$ Residual & 42.1 & 41.7 \\
\quad $+$ FFN & 34.9 & 36.1  \\
\toprule[1.0pt]
\end{tabular*}
\vspace{-1.0em}
\label{Ablation Study ASR Poor Performance}
\end{table}

\subsection{Discussion}

As demonstrated in Section~\ref{sec:asr_results}, independent Mamba or BiMamba models exhibit low performance in the ASR task, while replacing MHSA with BiMamba layers demonstrates impressive performance and outperforms the vanilla Transformer and Conformer models. Since the latter approach achieves significantly higher ASR performance by additionally employing a feed-forward network~(FFN) and a residual connection compared to the independent Mamba and BiMamba models, we then explore the factors contributing to this performance improvement via ablation studies in Table~\ref{Ablation Study ASR Poor Performance}. 

As illustrated in Section~\ref{sec:preliminary}, we consider a BiMamba layer as a weakly nonlinear module similar to MHSA~\cite{stanfordcs224n2024}. We thus assume that extracting high-abstraction-level information requires greater nonlinearity than capturing low-abstraction-level information. Specifically, ASR models transcribe speech signals by understanding the context of acoustic features and aligning speech to tokens, corresponding to low-level sequential and high-level semantic information, respectively. Since a speech enhancement model focuses on low-abstraction-level spectral information, an ASR model may need higher nonlinear ability than a speech enhancement model. This assumption aligns with the results presented in Table~\ref{ASR Main Result}, where replacing MHSA with Mamba modules in the Transformer does not consistently yield performance improvements, in contrast to their use within the Conformer model for ASR tasks. This indicates that the convolution modules in the Conformer provide more nonlinearities compared to the Transformer, enhancing the representation learning for ASR tasks.

\begin{figure}[!t]
    \centering
    \begin{subfigure}[b]{0.48\columnwidth} % 调整子图宽度为单列的一半
        \includegraphics[width=\textwidth]{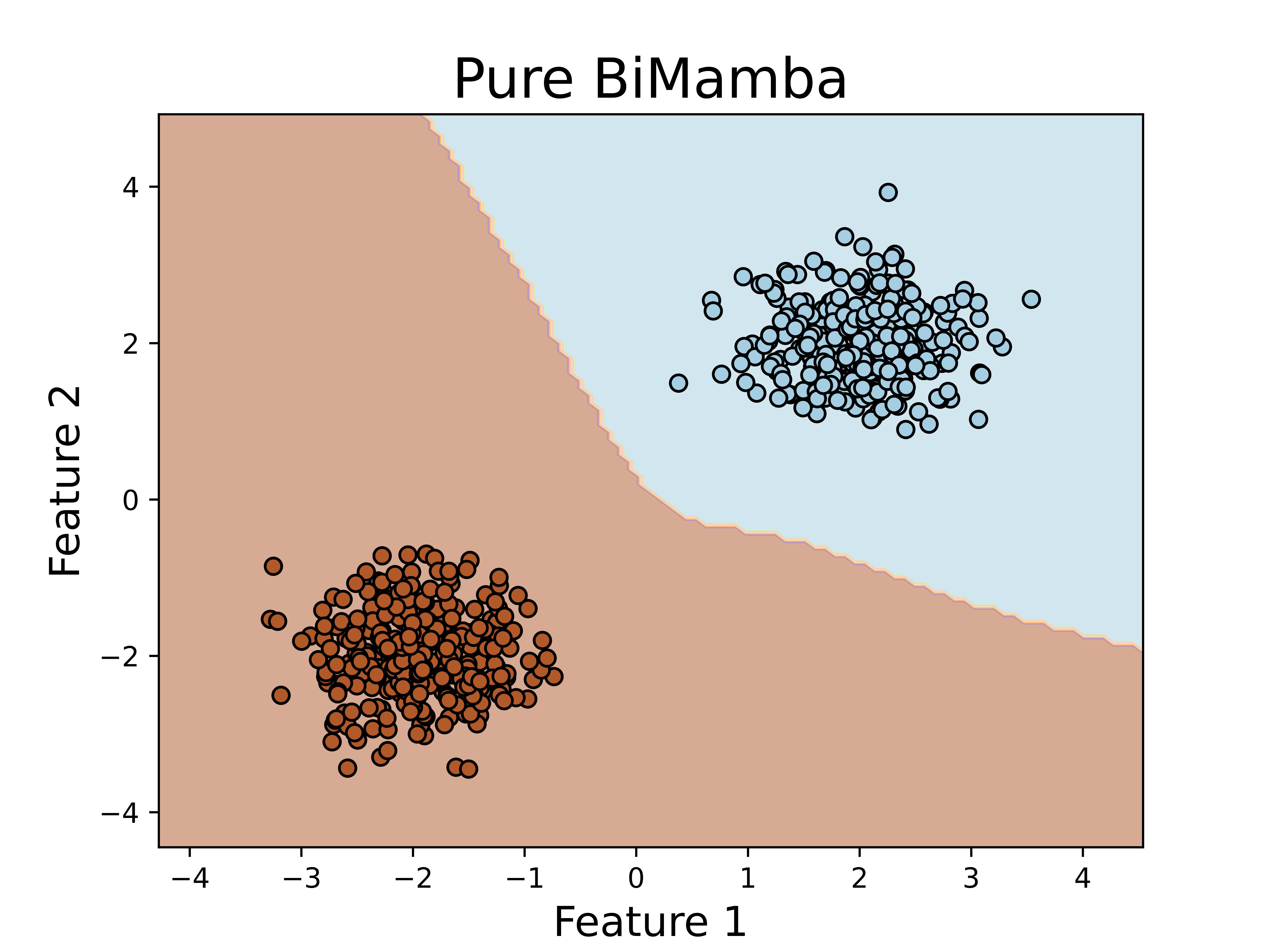}
        \caption{BiMamba}
        \label{Decision Boundary BiMamba}
    \end{subfigure}
    \hfill
    \begin{subfigure}[b]{0.48\columnwidth} % 调整子图宽度为单列的一半
        \includegraphics[width=\textwidth]{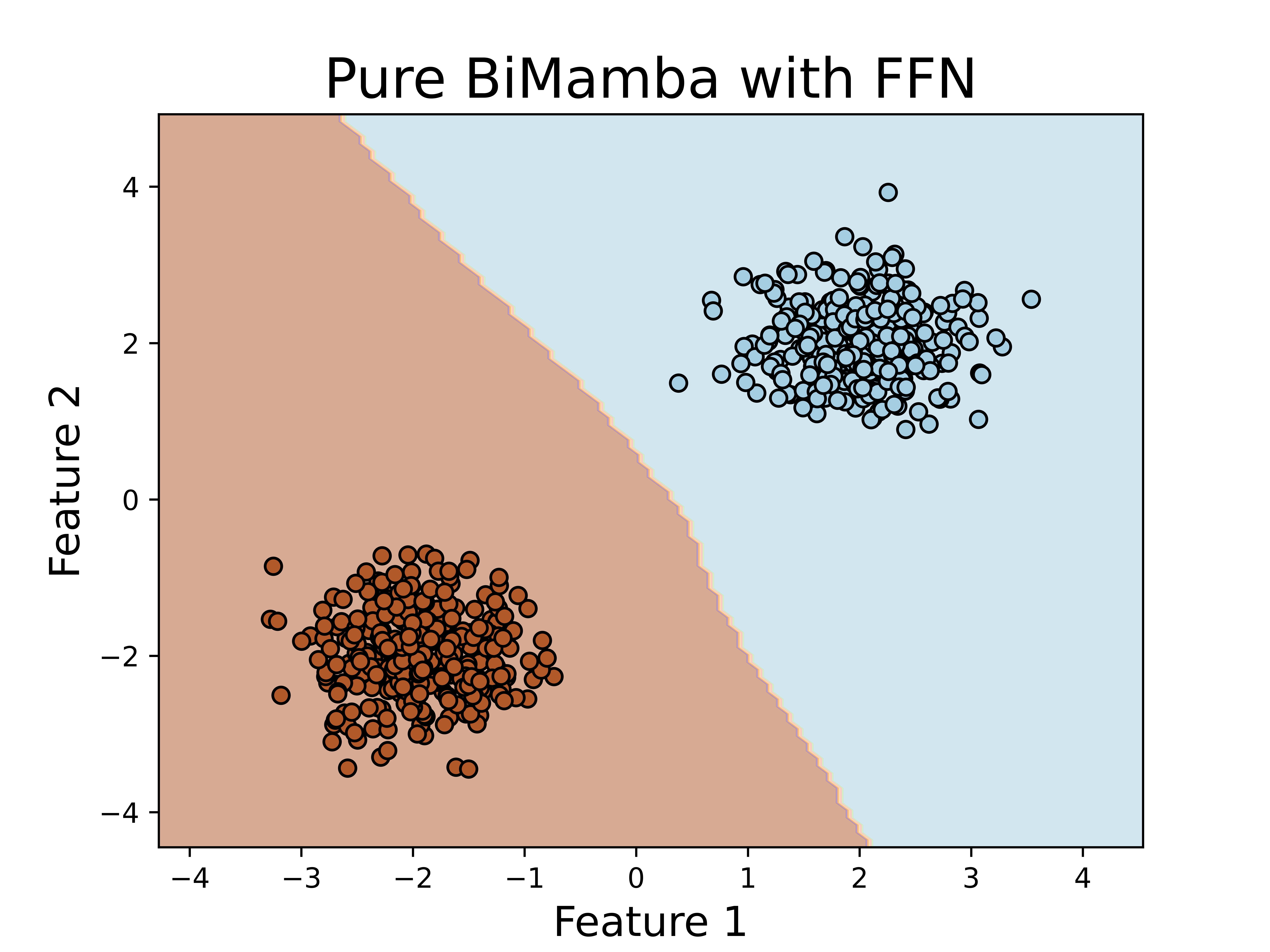}
        \caption{BiMamba with FFN}
        \label{Decision Boundary BiMamba FFN}
    \end{subfigure}
    \vfill
    \begin{subfigure}[b]{0.48\columnwidth} % 调整子图宽度为单列的一半
        \includegraphics[width=\textwidth]{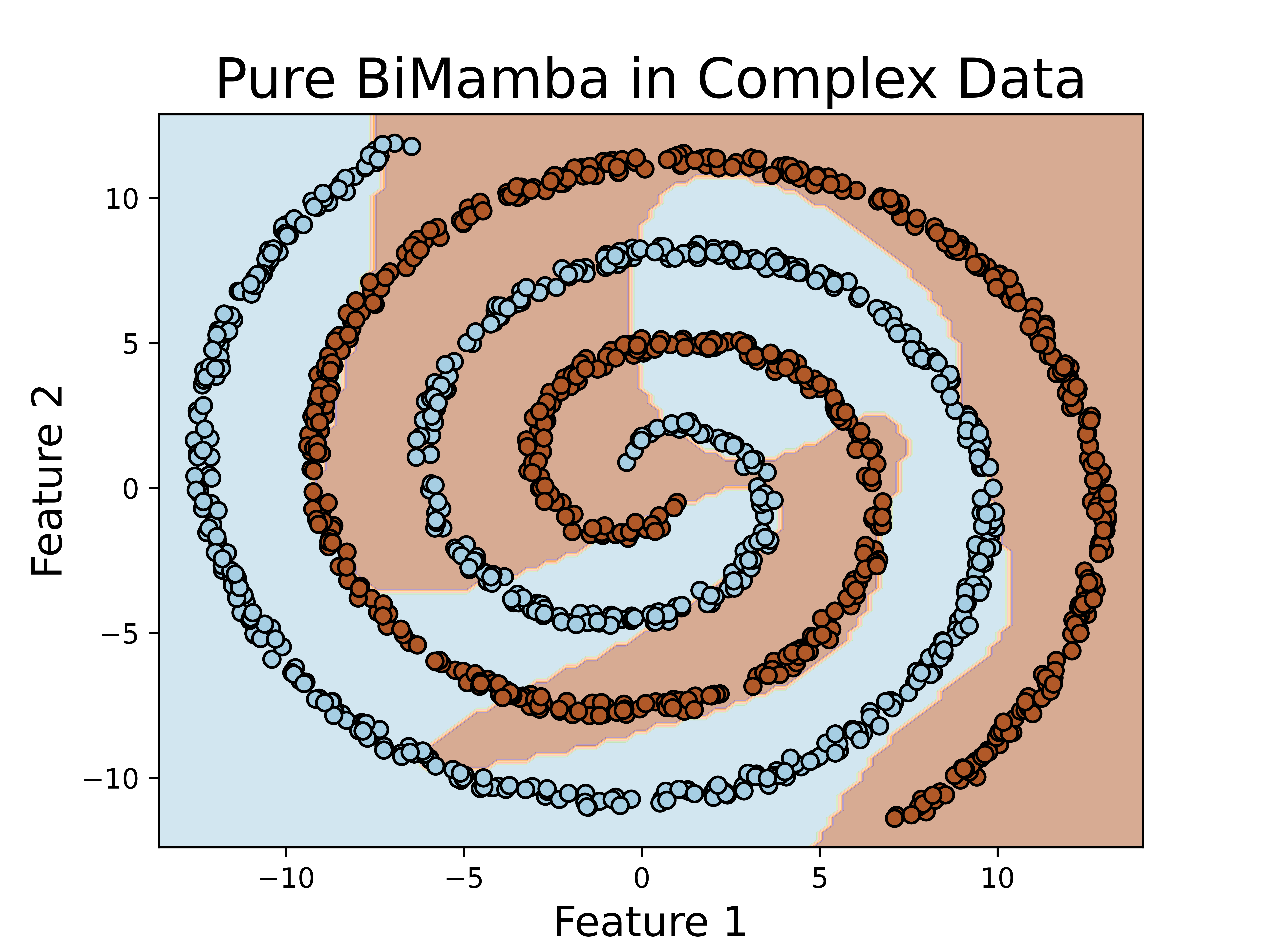}
        \caption{BiMamba}
        \label{Decision Boundary BiMamba complex}
    \end{subfigure}
    \hfill
    \begin{subfigure}[b]{0.48\columnwidth} % 调整子图宽度为单列的一半
        \includegraphics[width=\textwidth]{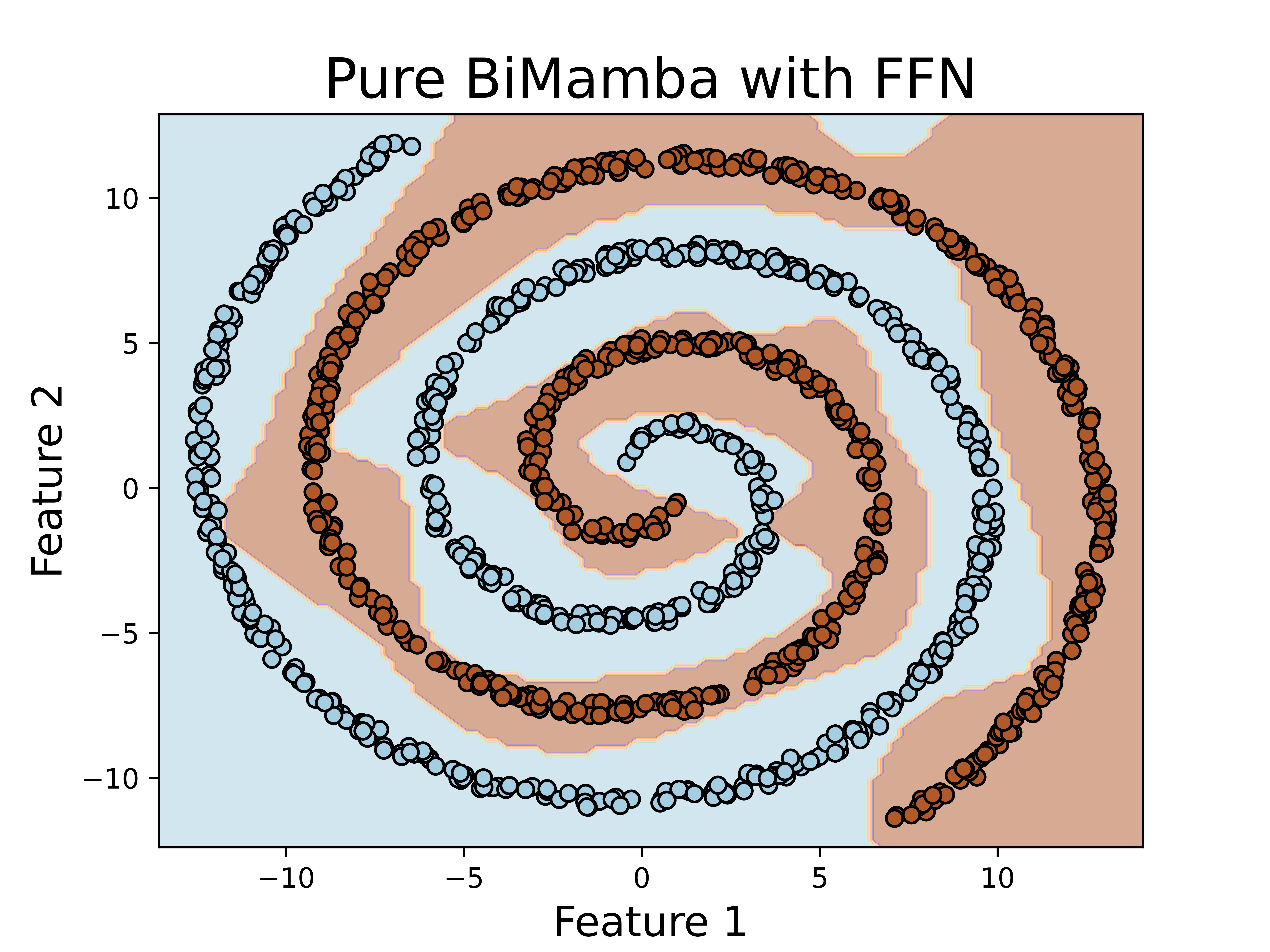}
        \caption{BiMamba with FFN}
        \label{Decision Boundary BiMamba FFN complex}
    \end{subfigure}
    \caption{Decision Boundaries for BiMamba and BiMamba with Feed-forward layer (FFN)}
    \vspace{-8pt}
    \label{fig:whole_db}
\end{figure}

To further validate this hypothesis, we design a visualization experiment using two synthetic datasets with different complexity levels: (1) a simple dataset containing two Gaussian clusters centered at (2,2) and (-2,-2) with a variance of 0.5; (2) a complex dataset consisting of two interleaved spirals with added Gaussian noise ($\text{noise level}=0.2$). We generate 800 samples for each dataset with an 80-20 train-test split. The models are trained using Adam optimizer with a learning rate of 0.01 for 100 epochs. The decision boundaries are visualized by evaluating model predictions on a 100×100 point grid spanning the input space. In Figure~\ref{fig:whole_db}, we use the BiMamba model to find the decision boundary for data with simple and complex distribution, respectively. We observe that BiMamba struggles to find the decision boundary for data of a complex distribution without the aid of an FFN (with ReLU activation similar to that in Transformer), which validates our assumption. 

In addition, results in Table~\ref{Ablation Study ASR Poor Performance} indicate the effectiveness of the FFN and residual connection in a Transformer model for ASR. Similar to independent ExtBiMamba, removing the residual connection and FFN from a Transformer model leads to gradient vanishing and decreases nonlinearity, resulting in significant performance degradation for ASR. This further underpins that using BiMamba layers as a replacement for MHSA is more appropriate for speech tasks which require models to learn high-abstraction-level information than employing it independently. Recent studies proposed that training stability of Mamba models benefits significantly from the application of normalization layers~\cite{fengempirical}, and suggested that the lack of inherent nonlinearity in Mamba necessitates the integration of components such as decoders or feed-forward layers to effectively handle the complexities of speech recognition~\cite{zhang2024rethinking}. These studies support our conclusion and extend it to training stability.

Similar to our investigation of Mamba in speech processing, numerous studies have explored different strategies for incorporating Mamba into ASR systems. Most of these works focus on integrating Mamba with other architectures (such as self-attention-based models)~\cite{jiang2024speech,chen2025selective,gao2024speech}. \cite{miyazaki2024exploring} took a different approach and demonstrated promising results with an independent Mamba encoder. Here, authors in~\cite{miyazaki2024exploring} observed that the Mamba model exhibited occasional instability during training. Therefore, fine-grained hyper-parameter tuning is required with an AdamW optimizer to achieve desirable performance. Our experiments are consistent with the observation that the Mamba model was sometimes unstable during training. To ensure fair comparisons between models, we directly transferred the optimal parameters from attention-based models to the corresponding experiments involving Mamba. We attempted to replicate the experiments following the setup in~\cite{miyazaki2024exploring} and found that our baseline performance improved by approximately 10\% when using the AdamW optimizer. However, we were unable to exactly reproduce their results, likely due to differences in the computational environment and the random seed. Since Mamba shows consistently high performance in ASR when used as a replacement for the self-attention module rather than independently, the above underpins our claim that using Mamba to replace the self-attention module is more appropriate for tasks requiring high-abstract-level information due to greater nonlinearity compared to independent Mamba.

Based on our findings and existing literature~\cite{miyazaki2024exploring, jiang2024speech}, we extend the scope of tasks that rely on low-abstraction-level information beyond speech enhancement to include phoneme recognition, speech separation, voice activity detection, and similar tasks. On the other hand, speech processing tasks that require high-abstraction-level information are not limited to speech recognition but also encompass speech translation, text-to-speech, and speech emotion recognition.

\section{Conclusion}
In this paper, we explore the use of Mamba in speech processing, for tasks requiring information from low to high abstraction levels. We first compared two bidirectional designs for Mamba and next employed them independently or as a replacement for MHSA in Transformer and Conformer models. While independent BiMamba models exhibited high performance in the speech enhancement task with the ability to capture low-abstraction-level spectral information, it can not well achieve speech recognition, which requires semantic information within the speech signal. In contrast, using BiMamba as a replacement of MHSA in Conformer (i.e., ConExtBiMamba) matched or exceeded the performance of the SoTA Branchformer across multiple datasets. Ablation studies suggest that using BiMamba to replace MHSA is more appropriate for tasks requiring high-abstract-level information due to greater nonlinearity compared to independent BiMamba.

% \section*{Acknowledgement}
% This work was supported by the Australian Research Council Discovery Project DP230101184.

\bibliographystyle{IEEEtran}
\bibliography{reference}

\end{document}